\documentclass[journal]{IEEEtran}
\usepackage{amsmath,amsfonts}
\usepackage{algorithmic}
\usepackage{array}
\usepackage{multirow}
\usepackage[caption=false,font=normalsize,labelfont=sf,textfont=sf]{subfig}
\usepackage{textcomp}
\usepackage{stfloats}
\usepackage{url}
\usepackage{cite}
\usepackage{verbatim}
\usepackage{graphicx}
\usepackage{amssymb}
\usepackage{colortbl}
\usepackage[table]{xcolor}
\usepackage[justification=centering]{caption}
\def\BibTeX{{\rm B\kern-.05em{\sc i\kern-.025em b}\kern-.08em
    T\kern-.1667em\lower.7ex\hbox{E}\kern-.125emX}}
\usepackage{balance}
    
\begin{document}
	
	\title{Content-Aware Quantization Index Modulation: Leveraging Data Statistics for Enhanced Image Watermarking}
	
	\author{Junlong Mao, Huiyi Tang, Shanxiang Lyu, Zhengchun Zhou, Xiaochun Cao
		\thanks{
		Junlong Mao, Huiyi Tang and Shanxiang Lyu are with the College of Cyber Security/College of Information Science and Technology, Jinan
		University, Guangzhou 510632, China (Emails: maojunlong@stu2022.jnu.edu.cn,  wt20180112@stu2019.jnu.edu.cn,
		 lsx07@jnu.edu.cn).
	  Zhengchun Zhou is with the School of Mathematics and the Key Laboratory
 	of Information Coding and Wireless Communications, Southwest Jiaotong
 		University, Chengdu 611756, China (E-mail: zzc@swjtu.edu.cn). 
 		Xiaochun Cao is with the School of Cyber Science and Technology, Sun Yat-sen University, Shenzhen Campus, Shenzhen 518107, China (E-mail: caoxiaochun@mail.sysu.edu.cn).
		}
		\thanks{This work has been submitted in part to 2023 IEEE Global Communications Conference (GLOBECOM) for possible presentation.}
	}

	
	\maketitle
	
	\begin{abstract}
		Image watermarking techniques have continuously evolved to address new challenges and incorporate advanced features. The advent of data-driven approaches has enabled the processing and analysis of large volumes of data, extracting valuable insights and patterns.
		In this paper, we propose two content-aware quantization index modulation (QIM) algorithms: Content-Aware QIM (CA-QIM) and Content-Aware Minimum Distortion QIM (CAMD-QIM). These algorithms aim to improve the embedding distortion of QIM-based watermarking schemes by considering the statistics of the cover signal vectors and messages. CA-QIM introduces a canonical labeling approach, where the closest coset to each cover vector is determined during the embedding process. An adjacency matrix is constructed to capture the relationships between the cover vectors and messages. CAMD-QIM extends the concept of minimum distortion (MD) principle to content-aware QIM. Instead of quantizing the carriers to lattice points, CAMD-QIM quantizes them to close points in the correct decoding region. Canonical labeling is also employed in CAMD-QIM to enhance its performance. 
		 Simulation results demonstrate the effectiveness of CA-QIM and CAMD-QIM in reducing embedding distortion compared to traditional QIM. The combination of canonical labeling and the minimum distortion principle proves to be powerful, minimizing the need for changes to most cover vectors/carriers. These content-aware QIM algorithms provide improved performance and robustness for watermarking applications.
	\end{abstract}
	
	
\section{Introduction}
\IEEEPARstart{D}{igital} watermarking is a technique used to embed information or digital markers into digital media, such as images, videos, audio files, or documents, without significantly altering the perceptual quality of the content \cite{cox2002digital,cayre2005watermarking}. This embedded information, known as a watermark, serves various purposes, including copyright protection, content authentication, and data integrity verification. It plays a crucial role in protecting the rights and integrity of digital media in an increasingly digital and interconnected world \cite{DBLP:journals/ieeemm/QinGZF23,DBLP:journals/sigpro/LefevreCFGH22,fernandez2022watermarking}.

Quantization index modulation (QIM) \cite{Chen2001} is a popular data hiding paradigm due to its considerable performance advantages over spread spectrum techniques. It excels in terms of information hiding capacity, perceptual transparency, robustness, and tamper resistance. QIM has been tailored to meet the demands of many applications. Some examples include angle QIM (AQIM) \cite{Ourique2005} and difference angle QIM (DAQIM) \cite{Cai2015a} for resisting gain attacks, dither-free QIM \cite{Lyu2023} to reduce the amount of synchronization information, $E_8$ lattice-based QIM \cite{Zhang2003} for enjoying the trade-off between computational complexity and robustness performance, the Tucker decomposition-based QIM \cite{Feng2016} for robust image watermarking, and minimum distortion QIM (MD-QIM) \cite{Lin2021} by moving the data point to the boundary of Voronoi regions to achieve smaller distortions.

Digital watermarking techniques have continuously evolved to meet the emerging challenges and threats in the field. One notable trend is the emergence of data-driven watermarking methods, which leverage advanced technologies such as machine learning and blockchain algorithms to generate and embed watermarks that are specifically tailored to the unique characteristics of the media content \cite{atli2022effectiveness,sahoo2020bdmark}. These data-driven approaches have shown promise in enhancing the robustness and imperceptibility of watermarking algorithms.
Advancements in technology have played a significant role in the development of robust and imperceptible watermarking algorithms. Researchers have explored various techniques and methodologies to improve the performance of watermarking systems. Robustness refers to the ability of a watermark to withstand attacks and intentional modifications, while imperceptibility refers to the extent to which the embedded watermark is perceptually invisible to human observers.

Several approaches have been proposed to enhance the robustness and imperceptibility of watermarking algorithms. Hybrid methods that combine multiple watermarking techniques, such as transform-based methods and spread spectrum techniques, have demonstrated improved performance in terms of both robustness and imperceptibility \cite{gong2021robust,singh2016hybrid}. These hybrid approaches leverage the strengths of different techniques to achieve a balance between robustness and imperceptibility.
Furthermore, comprehensive surveys and studies have been conducted to analyze the effectiveness of existing watermarking algorithms and identify areas for improvement \cite{agarwal2019survey}. These studies provide valuable insights into the strengths and limitations of different approaches and offer guidelines for developing more robust and imperceptible watermarking techniques.

However, despite the progress made in robust and imperceptible watermarking algorithms, there remains a need for further exploration of how data can be effectively exploited to enhance QIM and its variants for image watermarking. While data-driven watermarking techniques have shown promise in other domains, their application and effectiveness in the context of QIM-based watermarking are still relatively unexplored.

In this work, instead of relying on machine learning or blockchain, we propose a data-driven watermarking scheme by leveraging the statistics of data. Our goal is to enhance QIM or its variants by capturing the actual statistics of the data to be embedded. We employ lattices as a fundamental tool for theoretical analysis.

The contributions and highlights of this paper are as follows:

\begin{itemize}
	\item We introduce two novel approaches for image watermarking: Content-Aware QIM (CA-QIM) and Minimum-Distortion Content-Aware QIM (CAMD-QIM). These methods aim to achieve small-distortion watermarking while leveraging the statistical characteristics of the input messages and carriers.
	The core concept of our proposed approaches is to adaptively adjust the codebooks used for quantization in QIM based on the statistical properties of the input data. By doing so, we can effectively reduce the embedding distortion while maintaining robustness and applicability across different types of host signals and lattice bases.
	\item Exploiting tools from lattices and probability theory, we derive the mean square error (MSE) formulas of CA-QIM and CAMD-QIM. The derived results demonstrate that CA-QIM outperforms original QIM, while CAMD-QIM outperforms MD-QIM. We also derive the upper bound of the symbol error rate (SER) in additive white Gaussian noise (AWGN) channels. Our approach does not rely on the specific distribution of the host signals, thus showing positive effects on MSE, which is consistent with the simulation results.
	\item Numerical simulations justify the excellent performance of the proposed CA-QIM and CAMD-QIM. We adopt the discrete cosine transform (DCT) as the transformed domain for image processing. By embedding watermarks into the DCT domain, both of the proposed methods excel in terms of imperceptibility, embedding capacity, and SER. The simulation results demonstrate the merits of CA-QIM and CAMD-QIM, both of which outperform their counterparts.
\end{itemize}

This paper is organized as follows. Section \ref{Preliminaries} introduces the preliminaries of lattices and quantization index modulation. Section \ref{The Proposed Method} presents the proposed CA-QIM and CAMD-QIM. In Sections \ref{Distortion Analysis}, we analyze the distortion and SER of the two methods. Section \ref{Simulations} shows the simulation results. Conclusions are drawn in the final section.
	
	\section{Preliminaries}
	\label{Preliminaries}
This section provides a brief overview of the fundamental concepts and parameters related to lattices \cite{LyuWLC22,LyuWGHH21} and Quantization Index Modulation (QIM) \cite{Chen2001,Lyu2023}.
	\subsection{Lattices}
	An $N$-dimensional lattice $\Lambda$ in $\mathbb{R}^{N}$ is defined as a discrete additive subgroup $\Lambda = \left\lbrace \mathbf{Gz} | \mathbf{z} \in \mathbb{Z}^{N}\right\rbrace $, where $\mathbf{G}$ represents a set of $N$ linearly independent basic vectors $\mathbf{g}_{1}, \ldots, \mathbf{g}_{N}$ in $\mathbb{R}^{N}$. Figure 1 illustrates a three-dimensional lattice. The nearest neighbor quantizer $Q_\Lambda(\cdot)$ is defined as the function that maps a point $\mathbf{x}$ to its nearest lattice point:
	\begin{equation}
		Q_\Lambda(\mathbf{x}) = \mathop{\arg\min}_{\mathbf{\lambda} \in \Lambda} \left|  \mathbf{x} - \mathbf{\lambda} \right| .
	\end{equation}
	
	The Voronoi cell $\mathcal{V}_\lambda$ of a lattice point $\lambda \in \Lambda$ is the set of points which are closer to $\lambda$ than any other lattice points:
	\begin{equation}
		\mathcal{V}_\lambda = \{ \mathbf{x}: Q_\Lambda(\mathbf{x}) = \lambda \},
	\end{equation}
	and the fundamental Voronoi cell is the set of points which are quantized to the origin:
	\begin{equation}
		\mathcal{V}_\Lambda = \{ \mathbf{x}: Q_\Lambda(\mathbf{x}) = \mathbf{0} \}.
	\end{equation}
	The volume of the Voronoi cell  is defined by
	\begin{equation}
		\rm{Vol}(\mathcal{V}_\Lambda)=\int_{\mathcal{V}_{\Lambda}}d\mathbf{x}=|\rm{det}\mathbf{G}|.
	\end{equation}
	The normalized second moment of a lattice $\Lambda$ is defined by
	\begin{equation}
	\label{G_lambda}
		G(\Lambda)=\frac{1}{{\rm{Vol}}(\mathcal{V}_{\Lambda})^{\frac{2}{N}}} \times \frac{\int_{\mathbf{x}\in \mathcal{V}_{\Lambda} } ||\mathbf{x}||^2 \mathrm{d}\mathbf{x}}{N \rm{Vol}(\mathcal{V}_{\Lambda})}.
	\end{equation}
	Lattices naturally offer an elegant and efficient  way to pack the spheres in the Euclidean space. The packing spheres which do not intersect and the packing radius of a lattice is the inner radius of the Voronoi Cell $\mathcal{V}_0$. The packing radius $r_{pack(\Lambda)}$ is given by
	\begin{equation}
		r_{pack(\Lambda)}=\frac{1}{2}{\mathrm{d}}_{\min}(\Lambda),
	\end{equation}
	where $\mathrm{d}_{\min}(\Lambda)$ is the minimum distance between any two lattice points:
	\begin{equation}
		\mathrm{d}_{\min}(\Lambda) \triangleq \mathop{\min}_{\lambda \in \Lambda, \lambda \neq 0} \| \lambda \|.
	\end{equation}
	
	\begin{figure}[t]
		\centering
		\includegraphics[width=0.32\textwidth]{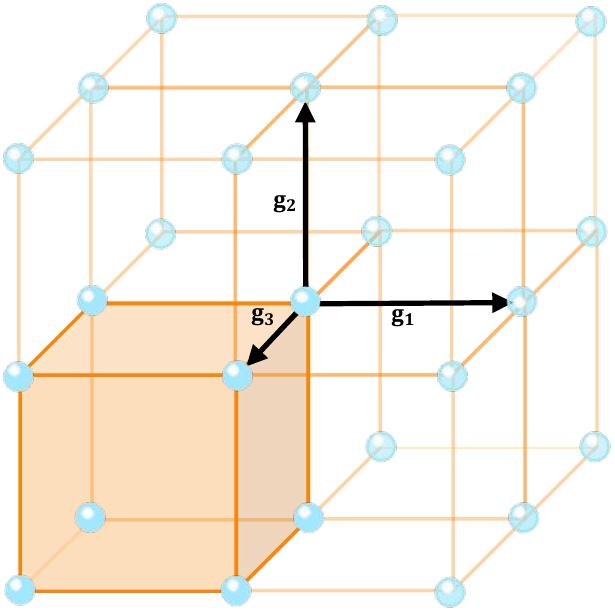}
		\caption{A three-dimensional lattice.}
		\label{lattice}
	\end{figure}
 
	Two $N$-dimensional lattices (${\Lambda}_c, {\Lambda}_f$) is nested if ${\Lambda}_c \subset {\Lambda}_f$. $\Lambda_f$ is called the fine lattice, and $\Lambda_c$ is called the coarse lattice. 
 Their  generator matrices are denoted by $\mathbf{G}_f$ and $\mathbf{G}_c$, satisfying
	\begin{equation}\label{eq_gene_matrix}
		\mathbf{G}_c=\mathbf{G}_f \cdot \mathbf{J},
	\end{equation}
	where the sub-sampling matrix $\mathbf{J}$ is an $n \times n$ integer matrix. The fine lattice $\Lambda_f$ can be decomposed as the union of $|\Lambda_f/\Lambda_c|$ cosets of the coarse lattice $\Lambda_c$:
	\begin{equation}\label{eq_setunion1}
		\Lambda_f 
		=\bigcup\limits_{{\mathbf{d}}_i \in \Lambda_f  {\slash} \Lambda_c} (\Lambda_c + {\mathbf{d}}_i),
	\end{equation}
	where each coset $\Lambda_c  + {\mathbf{d}}_i$ is a translated coarse lattice and ${\mathbf{d}}_i$ is called the coset representative of $\Lambda_c  + {\mathbf{d}}_i$. 
If $\mathbf{\Lambda}_c = \alpha \cdot \mathbf{\Lambda_f}$, then $|{\rm{det}}\mathbf{J}|={\alpha^N}$ and
 ${\Lambda}_c, {\Lambda}_f$ are called self-nested.
	
\subsection{Quantization Index Modulation (QIM) based Watermarking}

The QIM watermarking technique involves quantizing the cover signals into different lattice cosets, where each coset corresponds to a unique message. The process consists of two main phases: embedding and detection. 

In the embedding phase, the following steps are performed:

\noindent \textbf{i. [Lattice construction]} A pair of nested lattices $\Lambda_f$ and $\Lambda_c \subset \Lambda_f$ are considered, with their generators $\mathbf{G}_f$ and $\mathbf{G}_c$ related as specified in (\ref{eq_gene_matrix}).

\noindent \textbf{ii. [Labeling]} The message space $\mathcal{M}$ is defined as $\mathcal{M}=\mathbb{Z}^{N}_{\alpha}=\{0,\cdots,\alpha-1\}^{N}$. Given a host signal vector $\mathbf{s}$, to hide an information vector $\mathbf{m}_i \in \mathcal{M}$ within $\mathbf{s}$, a process called labeling is employed. One possible labeling solution is given by:
\begin{equation}
	\mathbf{d}_i = \mathcal{L}(\mathbf{m}_i) \triangleq \mathbf{G}_f \cdot \phi(\mathbf{m}_i),
\end{equation}
where $\phi : \mathbb{Z}^{N}_{\alpha} \rightarrow \mathbb{R}^{N}$ is a natural mapping function.

\noindent \textbf{iii. [Quantization]} The QIM encoder quantizes the host signal $\mathbf{s}$ to the nearest lattice point in $\Lambda_i$ using the following equation:
\begin{equation}
	\mathbf{s}_w = Q_{\Lambda_i}(\mathbf{s}) = Q_{\Lambda_c}(\mathbf{s} - \mathbf{d}_i) + \mathbf{d}_i.
\end{equation}
The index $i$ in $\Lambda_i$ indicates that the coset $\Lambda_i$ corresponds to the message $\mathbf{m}_i$, i.e., it is used to transmit the message $\mathbf{m}_i$. The payload of the embedding process is $\alpha^N$, resulting in a code rate per dimension of $R = \log \alpha$.

In the detection phase, the embedded watermark is extracted from the watermarked content to determine its presence and integrity. This allows authorized parties to verify the authenticity of the media and detect any unauthorized use or tampering. The detection algorithm of QIM can be described as follows: Assuming the received signal vector is $\mathbf{y} = \mathbf{x} + \mathbf{n}$, where $\mathbf{n}$ represents perturbation noise, the QIM technique performs the following de-quantization step to extract a coset representative:
\begin{equation}
	j = \mathop{\arg\min}_{i\in \left\lbrace 1, 2, \ldots, \alpha^N\right\rbrace} \mathrm{dist} (\mathbf{y}, \Lambda_i),
\end{equation}
where $\mathrm{dist}(\mathbf{y}, \Lambda_i) \triangleq \mathop{\min}_{\lambda \in \Lambda_i}\| \mathbf{y} - \lambda\|$. If the noise is sufficiently small such that $Q_{\Lambda_f}(\mathbf{n}) = 0$, then the embedded
message $\hat{\mathbf{m}}$ obtained from delabeling is correct. The estimated message is given by:
\begin{equation}
	\hat{\mathbf{m}} = \mathbf{G}_f^{-1} \cdot {\mathbf{d}_j}\ \mathrm{mod}\ \alpha.
\end{equation}

	\section{The Proposed Method}
	\label{The Proposed Method}
	In this section, we present the proposed method called Content-Aware Quantization Index Modulation (CA-QIM), which improves upon the labeling step of conventional QIM techniques by considering the statistics of the cover signal vector and the messages.
	
	\subsection{CA-QIM} 	\label{Proposed_CA-QIM}
	
	The CA-QIM method starts by finding the closest coset $\Lambda_i$ to each actual cover vector $\mathbf{s}_k$ using a similar process as the de-watermarking function:
	\begin{equation}
		\mathrm{Neighbor}(\mathbf{s}_k) = \mathop{\arg\min}_{i\in \left\lbrace 1, 2, \ldots, \alpha^N\right\rbrace} \text{dist} (\mathbf{y}, \Lambda_i).
	\end{equation}
	
	To incorporate the cover vector statistics and messages into the labeling process, an adjacency matrix $\mathbf{W}$ is constructed. The elements of $\mathbf{W}$ are initially set to zero. The matrix is updated by examining all pairs of $\mathbf{m}_i$ and $\mathbf{s}_k$. If $\mathrm{Neighbor}(\mathbf{s}_k) = j$ when embedding the information vector $\mathbf{m}_i$, then the corresponding element is incremented as follows:
	\begin{equation}
		w_{j,i} \rightarrow w_{j,i} + 1.
	\end{equation}
	Here, $w_{j,i}$ represents the element at row $j$ and column $i$ of $\mathbf{W}$. The resulting $\mathbf{W}$ is a square matrix with dimensions $\alpha^N \times \alpha^N$.
	
	In CA-QIM, the labeling step employs a canonical labeling approach that solves a maximum-weight matching problem. The problem can be formulated as follows:
	
	\textit{Problem: In a complete bipartite graph identified by the adjacency matrix}
	\begin{equation*}
    \mathbf{W}=
        \begin{bmatrix}
        \mathit{w}_{0,0}& \mathit{w}_{0,1} & \cdots & \mathit{w}_{0,\alpha^N-1}\\
        \mathit{w}_{1,0}& \mathit{w}_{1,1} & \cdots & \mathit{w}_{1,\alpha^N-1}\\
        \vdots &\vdots &\ddots &\vdots \\
        \mathit{w}_{\alpha^N-1,0}& \mathit{w}_{\alpha^N-1,1} & \cdots & \mathit{w}_{\alpha^N-1,\alpha^N-1}\\
        \end{bmatrix},
    \end{equation*}
\textit{find the matching $({\gamma_0,0}),({\gamma_1,1}), \ldots, (\gamma_{(\alpha^N-1)},{\alpha^N-1})$ such that the sum of weights}
\begin{equation}
	w_{\gamma_0,0}+w_{\gamma_1,1}+ \cdots w_{\gamma_{(\alpha^N-1)},\alpha^N-1}
\end{equation}
\textit{is maximized with the constraint $\gamma_0\neq \gamma_1 \neq \ldots \neq \gamma_{(\alpha^N-1)}$.}

The problem can be solved using the following procedure:

\begin{enumerate}
	\item Invert the sign of each element in $\mathbf{W}$.
	\item Subtract the minimum value from each row and column of the resulting matrix.
	\item Draw the minimum number of horizontal and vertical lines needed to cover all zeros in the matrix. If $\alpha^N$ lines are drawn, an optimal match can be found among the zeros from different rows and columns. Proceed to step 5. If fewer than $\alpha^N$ lines are drawn, proceed to step 4.
	\item Find the smallest uncovered element and subtract it from all uncovered elements. Add this value to all elements that are covered twice. Repeat from step 3.
\end{enumerate}

Based on the optimal matching $({\gamma_0,0}),({\gamma_1,1}), \ldots, (\gamma_{(\alpha^N-1)},{\alpha^N-1})$, the quantization function of CA-QIM is defined as:
\begin{equation}
	\mathbf{s}_w = Q_{\Lambda_{\gamma_i}}(\mathbf{s})
\end{equation}
for the message $\mathbf{m}_i$.

At the receiver's side, the estimated message is obtained by finding the coset representative index $\gamma_j$ that minimizes the distance between the received signal vector $\mathbf{y}$ and each coset $\Lambda_i$:
\begin{equation}
	\gamma_j =\mathop{\arg\min}_{i\in \left\lbrace 1, 2, \ldots, \alpha^N\right\rbrace} \text{dist} (\mathbf{y}, \Lambda_i).
\end{equation}
Finally, the estimated message is given by:
\begin{equation}
	\hat{\mathbf{m}}=\mathbf{G}_f^{-1} {\mathbf{d}_j}\ \mathrm{mod}\ \alpha.
\end{equation}

The CA-QIM method improves the conventional QIM technique by considering the cover vector statistics and messages during the labeling process, resulting in enhanced embedding performance. Two examples  are given below to demonstrate the rationale of CA-QIM.
	
	\begin{figure}[t]
		\centering
		\subfloat[]{
			\includegraphics[width=0.85\linewidth]{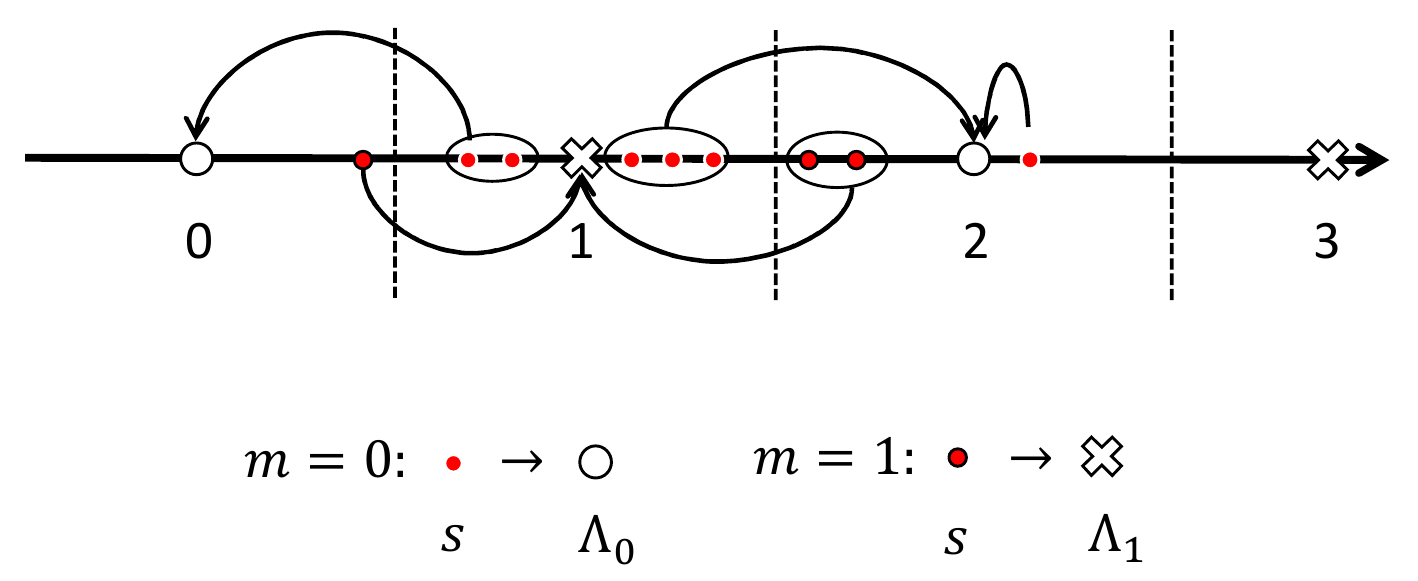}
			\label{lattice_Z_a}
		}
		
		\subfloat[]{
			\includegraphics[width=0.85\linewidth]{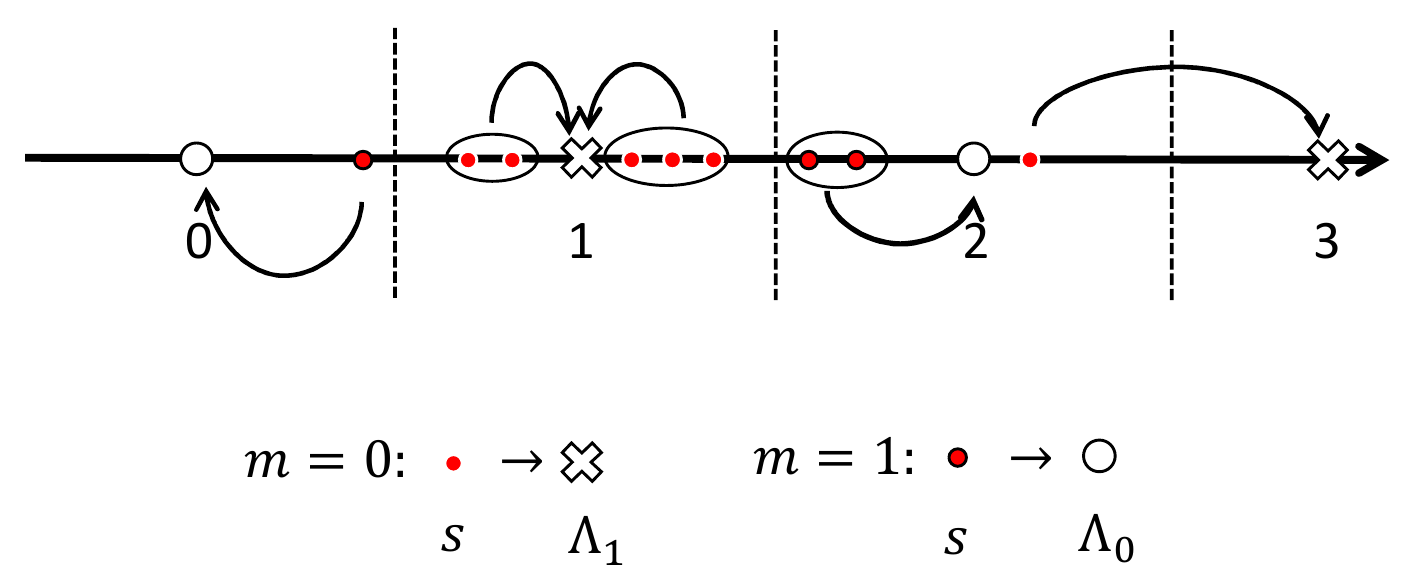}
			\label{lattice_Z_b}
		}
		\caption{The labeling schemes of QIM and CA-QIM. (a) QIM. (b) CA-QIM.}
		\label{lattice_Z}
	\end{figure}

	\emph{Example 1: } 
	Choose the typical 1-dimension QIM as comparison with CA-QIM, in which $\mathbb{Z}$ is ${\Lambda}_f$ and $2\mathbb{Z}$ is ${\Lambda}_c$. 
	The host signal $s \in \mathcal{S}$ is watermarked with a message $m \in \{0,1\}$ by
	\begin{equation}
	   s_w = Q_{{\Lambda}_i}(s)
	\end{equation}
	where
	\begin{equation}
\Lambda_0 = 2\mathbb{Z}, \, \Lambda_1 = 2\mathbb{Z}+1.
	\end{equation}
 As shown in Fig. \ref{lattice_Z_a}, the labeling scheme would lead to a quantization scheme that move $8$ of the $9$ carriers to ``far'' cosets ($|s_w - Q_{{\Lambda}_i}(s)|> 1/2$),
 and the adjacency matrix is 
 	\begin{equation*}
 	\mathbf{W}=
 	\begin{bmatrix}
 		1 & 3 \\
 		5 & 0
 	\end{bmatrix}.
 \end{equation*}
Regarding CA-QIM shown in Fig. \ref{lattice_Z_b}, it only moves $1$ carrier to a ``far'' coset. $8$ of the $9$ carriers satisfy $|s_w - Q_{{\Lambda}_i}(s)|< 1/2$.
  
	
	\emph{Example 2: } In terms of lattice construction, this example
	employs $D_2$ lattice as ${\Lambda}_f$, and the self-nested $2{\Lambda}_f$ as ${\Lambda}_c$.	
	The generator matrix of $D_2$ lattice is
	\begin{equation*}
		\mathbf{G}=
		\begin{pmatrix}
			1 & 0 \\[8pt]
			1 & 2\\
		\end{pmatrix}.
	\end{equation*}
	The 4 cosets are  given by 
	\begin{equation}
		\begin{aligned}
			\Lambda_0=\mathbf{G}([0,0]^\top+\mathbb{Z}^2),\nonumber\\
			\Lambda_1=\mathbf{G}([0,1]^\top+\mathbb{Z}^2),\nonumber\\
			\Lambda_2=\mathbf{G}([1,0]^\top+\mathbb{Z}^2),\nonumber\\
			\Lambda_3=\mathbf{G}([1,1]^\top+\mathbb{Z}^2).\nonumber\\
		\end{aligned}
	\end{equation}
	
    Assume that the cover vectors/carriers   are  \{(127,111), (35,120), (34,118), (89,37), (210,171), (145,101)\}. Then the indexes of their closet cosets,  $\mathrm{Neighbor}(\mathbf{s})$,  can be found:
    \{2, 3, 0, 2, 2, 1\}. Without loss of generality, set the probability of bit $0$ of the messages as  $0.9$. Then we have the indexes of these messages as \{0, 1, 0, 0, 0, 2\}. Thus the adjacency matrix of this example is
	\begin{equation*}
		\mathbf{W}=
		\begin{bmatrix}
			1 & 0 & 0 & 0\\
			0 & 0 & 1 & 0\\
			3 & 0 & 0 & 0\\
			0 & 1 & 0 & 0
		\end{bmatrix}.
	\end{equation*}
After 	solving the  maximum-weight matching problem, the optimal matching is given by  \{(2,0), (3,1), (1,2), (0,3)\}. The implication is that CA-QIM would quantize $\mathbf{s}$ via $\Lambda_2$ for $\mathbf{m}_0$, $\Lambda_3$ for $\mathbf{m}_1$,
$\Lambda_1$ for $\mathbf{m}_2$, and $\Lambda_0$ for $\mathbf{m}_3$. Thus in CA-QIM,  only $1$ carrier is
quantized to a ``far'' coset, while that number of QIM is $5$.
	
	
	
	\subsection{The minimum-distortion variant: CAMD-QIM}
	\label{Proposed_increase}
	A recent work by Lin et al. \cite{Lin2021} introduced a minimum distortion (MD) principle to reduce the embedding distortion of QIM, known as MD-QIM. In MD-QIM, the quantization step differs from conventional QIM in that the carriers are not required to be quantized to a lattice point. Instead, they are quantized to a nearby point within the correct decoding region. This trade-off sacrifices robustness to additive noises in favor of reduced embedding distortion. Inspired by this idea, we can apply the non-lattice quantization function of MD-QIM to construct a minimum-distortion content-aware QIM (CAMD-QIM) algorithm.
	
	Similar to CA-QIM, CAMD-QIM also employs canonical labeling. If a cover vector $\mathbf{s}$ lies within the Voronoi region of the fine lattice point $Q_{\Lambda_{\gamma_i}}(\mathbf{s})$, then $\mathbf{s}_w$ remains unchanged ($\mathbf{s}_w = \mathbf{s}$). However, if $\mathbf{s}$ falls outside the Voronoi region of $Q{\Lambda{\gamma_i}}(\mathbf{s})$, then $\mathbf{s}w$ is chosen as the closest point to $\mathbf{s}$ within the Voronoi region. This is achieved by the following equation:
	\begin{equation}
		\mathbf{s}_w = Q_{\Lambda_{\gamma_i}}(\mathbf{s}) - \frac{\mathbf{p}}{|\mathbf{p}|} \times (r_{\text{pack}(\Lambda_f)} - \epsilon),
	\end{equation}
	where $\epsilon$ is a small positive number used to move $\mathbf{s}_w$ away from the decision boundaries, and $\mathbf{p} = Q_{\Lambda_{\gamma_i}}(\mathbf{s}) - \mathbf{s}$. The detection algorithm for CAMD-QIM remains the same as that of CA-QIM.
	
\begin{figure}[t]
		\centering
		\subfloat[]{
			\includegraphics[width=0.85\linewidth]{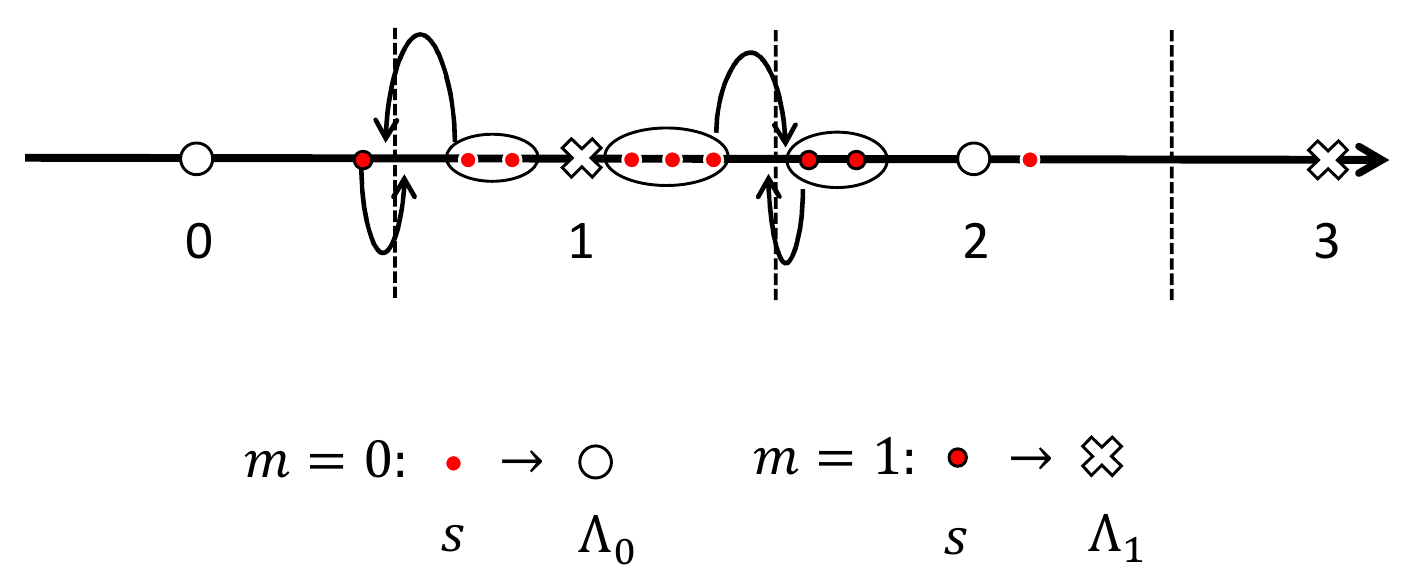}
			\label{lattice_Z_ae}
		}
		
		\subfloat[]{
			\includegraphics[width=0.85\linewidth]{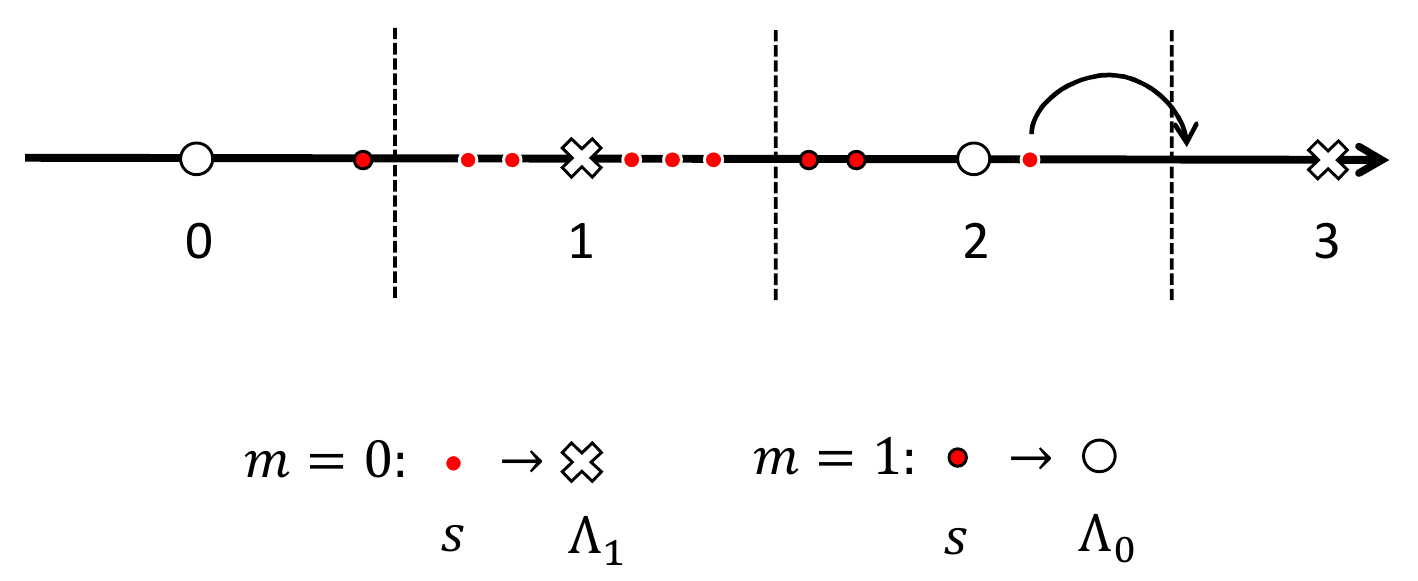}
			\label{lattice_Z_be}
		}
		\caption{Demonstration of MD-QIM and CAMD-QIM. (a) MD-QIM. (b) CAMD-QIM.}
		\label{lattice_Ze}
	\end{figure}

	In Figure \ref{lattice_Ze}, we present the rationale behind MD-QIM and CAMD-QIM using the same settings as Example 1. The figure illustrates that the combination of canonical labeling and the minimum distortion principle is highly effective, as most of the cover vectors/carriers do not require any changes.
	 
\section{Performance Analysis}
\label{Distortion Analysis}

Mean square error (MSE) is commonly used as a measure to quantify the difference between an estimator and the estimated values. In the context of watermarking algorithms, we can define the MSE to calculate the distortion as follows:

\begin{equation}
	\text{MSE} = \frac{1}{NM}\sum_{k=1}^{M} |\mathbf{s}_k - \mathbf{s}_{w,k}|^2,
\end{equation}
where $M$ is the total number of host vectors, $\mathbf{s}_k$ represents the host signal vectors before watermarking, and $\mathbf{s}_{w,k}$ represents the host signal vectors after watermarking.

The general distortion formula for QIM-alike algorithms can be expressed as:

\begin{equation}
	\text{MSE}{\text{QIM}} = \frac{1}{N} \int{\mathcal{V}_{\Lambda_c}} |\mathbf{x}|^2 \cdot f(\mathbf{x}) , d\mathbf{x},
\end{equation}
where $f(\mathbf{x})$ is the probability density function of the host signals, and $\mathcal{V}_{\Lambda_c}$ represents the Voronoi region of the coarse lattice. It should be noted that the above equation assumes uniformly distributed carriers over the Voronoi region of the coarse lattice, as described in \cite{Lin2021}. However, in this paper, we consider the case where the carriers are not necessarily uniformly distributed and proceed with the analysis accordingly.
	
	\subsection{Distortion of CA-QIM}
	Without loss of generality, denote the probability of message $\mathbf{m}_i$ as $p_i$, the probability of  $\mathbf{s}_k$ as $q_k$. The probability of quantizing $\mathbf{m}_i$ into the coset $\Lambda_k$ of $\mathbf{s}_k$ is $p_i\cdot q_k$. After applying CA-QIM, probabilities get sorted such that $p_{i_1} \leq p_{i_2} \leq \cdots \leq p_{i_{\alpha^N}}$ and $q_{k_1} \leq q_{k_2} \leq \cdots \leq q_{k_{\alpha^N}}$, and host signals and messages with same order of probabilities are regarded as remapped pairs as proposed in \ref{Proposed_CA-QIM}.

    According to the flat-host assumption, we can analyze the probability of whether the carriers are in the corresponding coset or not, i.e., 
	\begin{equation}{\label{P1}}
		P_1=P(\mathbf{m} \in \mathcal{V}_{\Lambda_f}) = \sum_{j=1}^{\alpha^N} p_{i_j} q_{k_j},
	\end{equation}
	\begin{equation}{\label{P2}}
		P_2
		=P(\mathbf{m}\in \mathcal{V}_{\Lambda_c} \backslash \mathcal{V}_{\Lambda_f}) 
		= 1 - \sum_{j=1}^{\alpha^N} p_{i_j} q_{k_j}.
	\end{equation}
	Correspondingly, the overall expected MSE of CA-QIM can be formulated as
	\begin{equation}
	    \begin{aligned}
	    &MSE_{CA\text{-}QIM} = \\&\frac{1}{N} \Bigg[P_1 \int_{\mathcal{V}_{\Lambda_f}}\|\mathbf{x}\|^2 d\mathbf{x} + P_2 \int_{\mathcal{V}_{\Lambda_c} \backslash \mathcal{V}_{\Lambda_f}}\|\mathbf{x}\|^2 d\mathbf{x}\Bigg].
	\end{aligned}
	\end{equation}

	For the original QIM, the probabilities corresponding to the above two cases respectively are
	\begin{equation}
		P_1'=P(\mathbf{m} \in \mathcal{V}_{\Lambda_f}) = \sum_{j=1}^{\alpha^N} p_{j} q_{j},
	\end{equation}
	\begin{equation}
		P_2'=P(\mathbf{m}\in \mathcal{V}_{\Lambda_c} \backslash V_{\Lambda_f}) = 1 - \sum_{j=1}^{\alpha^N} p_{j} q_{j}.
	\end{equation}
	Obviously, $P_1 \geqslant P_1'$ and $P_2 \leq P_2'$. And the overall expected MSE of QIM can be calculated as
	\begin{equation}
	\label{MSE_QIM}
	    \begin{aligned}
        &MSE_{QIM} 
        \\&= \mathbb{E}\big(\frac{\int_{\mathcal{V}_{\Lambda_c}}\|\mathbf{x}\|^2 \cdot f(\mathbf{x})d\mathbf{x}}{N}\big)
        \\&= \frac{1}{N} \Bigg[P_1' \int_{\mathcal{V}_{\Lambda_f}}\|\mathbf{x}\|^2 d\mathbf{x} + P_2' \int_{\mathcal{V}_{\Lambda_c} \backslash \mathcal{V}_{\Lambda_f}}\|\mathbf{x}\|^2 d\mathbf{x}\Bigg].
        \end{aligned}
	\end{equation}
  If the host signals and messages are modeled as uniform distribution, then $P_1 = P_1'= \sum_{j=1}^{\alpha^N} \frac{1}{{\alpha^N}^2} = \frac{1}{{\alpha^N}}$ and $P_2 = P_2' = 1-\frac{1}{{\alpha^N}}$. In this case the MSE becomes
    \begin{equation}
        \begin{aligned}
            MSE_{CA\text{-}QIM} &= MSE_{QIM} \\&= \frac{1}{N} \frac{1}{\rm{Vol}(\mathcal{V}_{\Lambda_c})}\int_{\mathcal{V}_{\Lambda_c}}\|\mathbf{x}\|^2d\mathbf{x}
           \\&=G(\Lambda_c)\rm{Vol}(\mathcal{V}_{\Lambda_c})^{\frac{2}{N}},
        \end{aligned}
    \end{equation}
	which indicates that CA-QIM does perceive the content of signals and messages.
	
	\subsection{Distortion of CAMD-QIM}
	MD is applied in the process of quantizing after labeling. We use $MSE_{CAMD\text{-}QIM}'$ and $MSE_{CAMD\text{-}QIM}''$ to represent the MSE of CAMD-QIM in the case of (\ref{P1}) and (\ref{P2}), respectively.
	
\begin{figure}[th]
		\centering
		\subfloat[]{\includegraphics[width=0.27\linewidth]{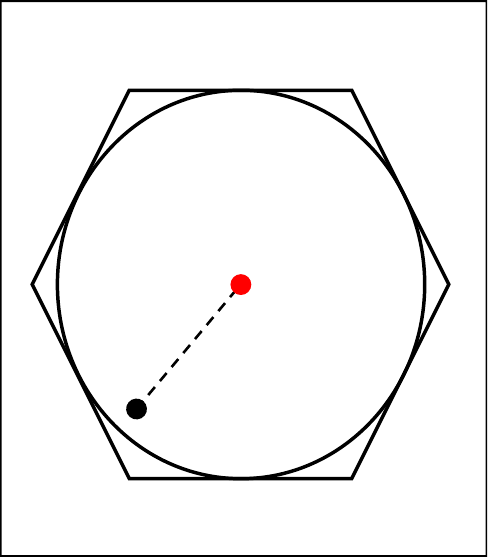}\label{fig3b}}
		\subfloat[]{\includegraphics[width=0.419\linewidth]{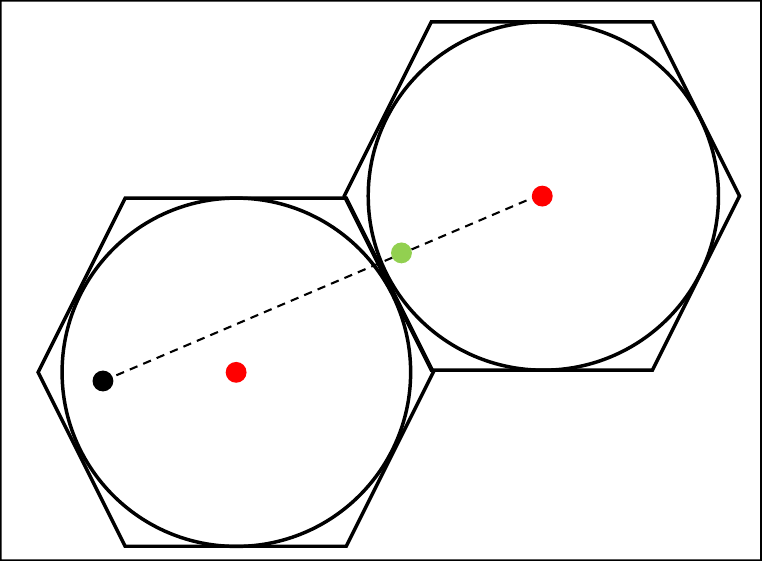}\label{fig3c}}
		\caption{The example over an $A_2$ lattice. (a) Host signal vector in the corresponding Voronoi region. (b) Host signal vector in incongruous Voronoi region.}
		\label{Packing sphere}
	\end{figure}

	In the case of $\mathbf{s}\in \mathcal{V}_{\Lambda_f}$ as shown  in  Fig.\ref{Packing sphere}\subref{fig3b}, the host signal is in the corresponding Voronoi region. From \ref{Proposed_increase}, the carriers remain  in the initial positions. So the MSE of CAMD-QIM is
	\begin{equation}
	\label{MSE_CAMD'}
	    MSE_{CAMD\text{-}QIM}' = 0.
	\end{equation}
	
	In the case of $\mathbf{s} \in \mathcal{V}_{\Lambda_c} \backslash \mathcal{V}_{\Lambda_f}$, the host signal needs to be moved into the packing sphere of corresponding Voronoi region. As depicted in Fig.\ref{Packing sphere}\subref{fig3c}, the host signal is in incongruous Voronoi region.  The MSE in this scenario can be calculated as
	\begin{equation}
		\label{MSE_CAMD''}
		\begin{aligned}
			MSE_{CAMD\text{-}QIM}''
			&=\frac{1}{N}\int _{\mathcal{V}_{\Lambda_c} \backslash \mathcal{V}_{\Lambda_f}}\|\mathbf{x}-(r_{pack(\Lambda_f)}-\epsilon)\|^2d\mathbf{x}.
		\end{aligned}
	\end{equation}

	Then based on (\ref{MSE_CAMD'}) and (\ref{MSE_CAMD''}), the overall expected MSE of CAMD-QIM is
	\begin{equation}
	\label{MSE_CAMD-QIM}
	    \begin{aligned}
	        &MSE_{CAMD\text{-}QIM}\\
	        &=P_1 \cdot MSE_{CAMD\text{-}QIM}' + P_2 \cdot  MSE_{CAMD\text{-}QIM}''\\
	        &=\frac{(1-\sum_{j=1}^{\alpha^N} p_{i_j} q_{k_j})}{N}\Bigg[\int_{\mathcal{V}_{\Lambda_c} \backslash \mathcal{V}_{\Lambda_f}}\|\mathbf{x}\|^2d\mathbf{x}\\
	        &-2(r_{pack(\Lambda_\mathit{f})}-\epsilon)\int_{\mathcal{V}_{\Lambda_c} \backslash \mathcal{V}_{\Lambda_\mathit{f}}}\|\mathbf{x}\|d\mathbf{x}
	        + \rm{Vol}(\mathcal{V}_{\Lambda_c} \backslash \mathcal{V}_{\Lambda_\mathit{f}}) \Bigg].
	    \end{aligned}
	\end{equation}
	With reference to  (\ref{G_lambda}), we have
	\begin{equation}
	\label{||x||^2}
	    \int _{\mathcal{V}_{\Lambda_c} \backslash \mathcal{V}_{\Lambda_\mathit{f}}}\|\mathbf{x}\|^2d\mathbf{x} = N G(\Lambda_c \backslash \Lambda_\mathit{f}) \rm{Vol}(\mathcal{V}_{\Lambda_c} \backslash \mathcal{V}_{\Lambda_\mathit{f}})^{\frac{2}{N}+1}.
	\end{equation}
	Through using Cauchy-Schwarz inequality, we have
	\begin{equation}
	\label{Cauchy-Schwarz}
	    \begin{aligned}
	        &\int _{\mathcal{V}_{\Lambda_c} \backslash \mathcal{V}_{\Lambda_\mathit{f}}}\|\mathbf{x}\|d\mathbf{x}\leq
	        \sqrt{\int _{\mathcal{V}_{\Lambda_c} \backslash \mathcal{V}_{\Lambda_\mathit{f}}}\|\mathbf{x}\|^2d\mathbf{x}}
	        \times\sqrt{\rm{Vol}(\mathcal{V}_{\Lambda_c} \backslash \mathcal{V}_{\Lambda_\mathit{f}})}
	        \\&\leq\sqrt{NG(\Lambda_c)\rm{Vol}(\mathcal{V}_{\Lambda_c})^{\frac{2}{N}+2} - NG(\Lambda_\mathit{f})\rm{Vol}(\mathcal{V}_{\Lambda_\mathit{f}})^{\frac{2}{N}+2}}.
	    \end{aligned}
	\end{equation}
	
Combining (\ref{MSE_CAMD-QIM}), (\ref{||x||^2}) and (\ref{Cauchy-Schwarz}), the MSE of CAMD-QIM has the upper bound of
	\begin{equation}
	\label{MSE_CAMD-QIM_upper}
	    \begin{aligned}
	        &MSE_{CAMD\text{-}QIM} \\
	        &\leq\frac{(1-\sum_{j=1}^{\alpha^N} p_{i_j} q_{k_j})}{N}
	        \Bigg[N G(\Lambda_c) \rm{Vol}(\mathcal{V}_{\Lambda_c})^{\frac{2}{N}+1} \\
	        &- N G(\Lambda_\mathit{f}) \rm{Vol}(\mathcal{V}_{\Lambda_\mathit{f}})^{\frac{2}{N}+1} - 2(r_{pack(\Lambda_\mathit{f})}-\epsilon)\\
	        &\cdot\sqrt{N G(\Lambda_c) \rm{Vol}(\mathcal{V}_{\Lambda_c})^{\frac{2}{N}+2} - N G(\Lambda_\mathit{f})
	        \rm{Vol}(\mathcal{V}_{\Lambda_\mathit{f}})^{\frac{2}{N}+2}}\\
	        &+ \rm{Vol}(\mathcal{V}_{\Lambda_c}) - \rm{Vol}(\mathcal{V}_{\Lambda_\mathit{f}}) \Bigg].
	    \end{aligned}
	\end{equation}
	It can be observed from the above formula that when the lattice construction is fixed, the only parameter that determines the MSE is the probability of mapping between carriers and messages. If $P_2=0$, the MSE of CAMD-QIM is $0$.

	\subsection{Spatial-Frequency Domain Distortion}
	\label{Image Transform}
The relationship between mean square error (MSE) in the spatial domain and frequency domain is examined in this section, specifically focusing on the discrete cosine transform (DCT) domain. The $8\times 8$ block DCT transform is defined by the following equation:
	\begin{equation}
	    \begin{aligned}
	        \mathbf{Y}(u,v)&=\frac{1}{4}C(u)C(v)\\
	        &\cdot\sum_{m=0}^{7}\sum_{n=0}^{7}\mathbf{X}(m,n)\cos{\frac{(2m+1)u\pi}{16}}\cos{\frac{(2n+1)v\pi}{16}}
	    \end{aligned}
	\end{equation}
	where $\mathbf{X}(m,n)$ represents the pixel values of the original $8\times 8$ image block,
	\begin{equation*}
	    C(u)= \begin{cases}
              1/\sqrt{2},\quad &u=0, \\
              1,\quad & \rm{others},
              \end{cases} 
	\end{equation*}
	and 
	\begin{equation*}
	    C(v)= \begin{cases}
              1/\sqrt{2},\quad &v=0, \\
              1,\quad & \rm{others}.
              \end{cases} 
	\end{equation*}

	\begin{figure*}
	\begin{equation}
		\label{transform_matrix}
		\mathbf{T} = \left[
		\begin{array}{cccccccc}
			\Gamma(4) & \Gamma(4) & \Gamma(4) & \Gamma(4) & 
			\Gamma(4) & \Gamma(4) & \Gamma(4) & \Gamma(4) \\
			\Gamma(1) & \Gamma(3) & \Gamma(5) & \Gamma(7) & 
			-\Gamma(7) & -\Gamma(5) & -\Gamma(3) & -\Gamma(1) \\
			\Gamma(2) & \Gamma(6) & -\Gamma(6) & -\Gamma(2) & 
			-\Gamma(2) & -\Gamma(6) & \Gamma(6) & \Gamma(2) \\
			\Gamma(3) & -\Gamma(7) & -\Gamma(1) & -\Gamma(5) & 
			\Gamma(5) & \Gamma(1) & \Gamma(7) & -\Gamma(3) \\
			\Gamma(4) & -\Gamma(4) & -\Gamma(4) & \Gamma(4) & 
			\Gamma(4) & -\Gamma(4) & -\Gamma(4) & \Gamma(4) \\
			\Gamma(5) & -\Gamma(1) & \Gamma(7) & \Gamma(3) & 
			-\Gamma(3) & -\Gamma(7) & \Gamma(1) & -\Gamma(5) \\
			\Gamma(6) & -\Gamma(2) & \Gamma(2) & -\Gamma(6) & 
			-\Gamma(6) & \Gamma(2) & -\Gamma(2) & \Gamma(6) \\
			\Gamma(7) & -\Gamma(5) & \Gamma(3) & -\Gamma(1) & 
			\Gamma(1) & -\Gamma(3) & \Gamma(5) & -\Gamma(7) 
		\end{array}
		\right]
	\end{equation}
	\rule[-5pt]{18.07cm}{0.1em}
\end{figure*}

The transformation matrix for the $8\times 8$ block DCT is defined using the concept that the $K$-point DCT of any sequence $\mathbf{s}$ can be seen as a projection onto an orthogonal basis \cite{feig1992fast}. In this case, the transformation matrix is given by Equation (\ref{transform_matrix}), where $\Gamma(k)=\cos{({k\pi}/{16})}$. 
	Thus, the transform of $8\times 8$ block DCT is formulated as
	\begin{equation}
	    \mathbf{Y} = \mathbf{T} \mathbf{X} \mathbf{T}^{\rm{T}},
	\end{equation}
	where $\mathbf{X}$ and $\mathbf{Y}$ respectively represent an $8\times 8$ image block of spatial domain and DCT domain. 
	
	 Assume that $\mathbf{Y}$ has gotten distortions to become $\mathbf{Y'}$. Then with $\Delta \mathbf{Y} = \mathbf{Y} - \mathbf{Y'}$,  $\Delta \mathbf{X} = \mathbf{X} - \mathbf{X'}$, we have $\Delta \mathbf{Y} = \mathbf{T} \Delta \mathbf{X} \mathbf{T}$. According to the properties of Euclidean norms, the MSE over carrier $\mathbf{X}$ and $\mathbf{Y}$ can be written respectively as
	 \begin{equation}
	     MSE_\mathbf{X} = \frac{1}{64} \|\Delta \mathbf{X}\|^2_2,
	 \end{equation}
	 and 
	 \begin{equation}
	     MSE_\mathbf{Y} = \frac{1}{64} \|\Delta \mathbf{Y}\|^2_2 = \frac{1}{64} \|\mathbf{T} \Delta \mathbf{X} \mathbf{T}^{\rm{T}}\|^2_2.
	 \end{equation}
	 
	 From (\ref{transform_matrix}), it can be verified that $\mathbf{T}$ is an orthogonal matrix, and $\|\mathbf{T}\|^2_2 = 1$. Then it follows from the property of norms that
	 \begin{equation}
	    \begin{aligned}
	        MSE_\mathbf{Y} &= \frac{1}{64} \|\mathbf{T} \Delta \mathbf{X} \mathbf{T}^{\rm{T}}\|^2_2 \\
	        & = \frac{1}{64} \|\mathbf{T}\|^2_2  \|\Delta \mathbf{X}\|^2_2 \|\mathbf{T}^{\rm{T}}\|^2_2 \\
	        & = \frac{1}{64} \|\Delta \mathbf{X}\|^2_2 = MSE_\mathbf{X},
	    \end{aligned}
	 \end{equation}
	 which shows that with distortion, images in the spatial domain and frequency domain are affected to the same extent. If the stored    pixels of the images can only be of integer formats, we have $MSE_\mathbf{X} \leq  MSE_\mathbf{Y}$ in general.
	 
	 \subsection{Symbol Error Rate Analysis}
	 \label{Symbol Error Rate Analysis}
	 In the proposed schemes, CA-QIM has noise tolerance ability while CAMD-QIM does not. So this section only analyzes the detection error rate performance of CA-QIM. Considering
	 the watermarked carriers $\mathbf{s}_w$ may go through malicious additive noise attacks or oblivious addition noise pollution, the received signal vector at the receiver's side is $\mathbf{y} = \mathbf{s}_w + \mathbf{n}$. Since the lattice is geometrically uniform, the point error probability has the upper bound \cite{boutros1996good}
	 \begin{align}
	 	\label{point error}
	 	P_e(\Lambda_f) \leq \sum_{\mathbf{z}\ne\mathbf{s}_w, \mathbf{z}\in \Lambda_f}P_e(\mathbf{s}_w\to\mathbf{z}),
	 \end{align}
	 where $P_e(\mathbf{s}_w\to\mathbf{z})$ is the   probability that $\mathbf{y}$ lies in the Voronoi region centered at $\mathbf{z}$ while transmitting $\mathbf{s}_w$.
	 
	 In the AWGN channel, (\ref{point error}) becomes
	 \begin{align}
	 	P_e(\Lambda_f) \leq \frac{\tau}{2} \mathrm{erfc}(\frac{d_\mathrm{min}/2}{\sqrt{2N_0}}),
	 \end{align}
	 where $\tau$ is the kissing number, $d_\mathrm{min}$ is the minimum Euclidean distance of the lattice and $N_0$ is the noise power spectral density. Thus the vector error rate in the transmission model can be regarded as the  probability of $\mathbf{y}$ lying outside the Voronoi region  $\mathcal{V}_{\Lambda_f}$ centered at $\mathbf{s}_w$. Thus we have the following upper bound for the vector error rate:
	 \begin{align}
	 	P(\mathbf{y} \notin \mathcal{V}_{\Lambda_f}) \leq  P_e(\Lambda_f).
	 \end{align}
	 By averaging over the received signal vectors $\mathbf{y}_j$, the symbol error rate (SER) of CA-QIM is given by
	 \begin{align}
	 	SER = \frac{\sum_{j=1}^{M} P(\mathbf{y}_j \notin \mathcal{V}_{\Lambda_f})}{M} \leq \frac{\tau}{2} \mathrm{erfc}(\frac{d_\mathrm{min}/2}{\sqrt{2N_0}}),
	 \end{align}
	 where $M$ is the total number of host vectors.

	\section{Simulations}
	\label{Simulations}
	In this section, we conduct numerical simulations to verify the effectiveness of the proposed CA-QIM and CAMD-QIM.
	
	\subsection{Setups}

We utilized two standard image databases for conducting watermarking experiments. Set12 \cite{zhang2017beyond} comprises grayscale images of diverse scenes, with each image's size being either 256×256 or 512×512 pixels. Another database, BSD68 \cite{martin2001database}, consists of 68 grayscale images that vary in size.   The AC coefficients of images subjected to discrete cosine transform (DCT) follow an approximate Laplacian distribution \cite{joshi1995comparison}. To simulate the common JPEG image compression technique, we divide the image into non-overlapping blocks of size 8×8 pixels. Since JPEG compression employs DCT, embedding messages into the frequency domain aligns with this process. The low, mid, and high-frequency coefficients correspond to the front, middle, and rear parts of the zig-zag-scan sequence within each block. Different frequencies of each block can accommodate single or multiple messages.
	
To ensure comprehensive evaluation, we utilize $A_2, D_4$, and $E_8$ lattices as the fine lattices (denoted as $\Lambda_f$). The corresponding coarse lattices for these fine lattices are $\alpha \Lambda_f$, where $\alpha$ represents the parameter governing the code rate of messages. We reference benchmark algorithms from \cite{Moulin2005} and \cite{Lin2021}, collectively referred to as QIM, for comparison. Additionally, we implement an enhanced version of the MD-QIM algorithm to evaluate against our proposed schemes.
	
On one hand, according to the definition of the code rate, as the code rate increases, the distortion also increases. On the other hand, since image pixels are rounded to integers, it is necessary to set a larger code rate to ensure that the Voronoi region includes multiple integers. Therefore, we set the code rate as $R=2$ and maintain a probability of $0.9$ for the occurrence of bit $0$ in the message, aligning with the hypothesis proposed in this paper.
	
	\subsection{Imperceptibility measurement}

To ensure a comprehensive quantitative evaluation, we incorporate additional metrics alongside Mean Squared Error (MSE). One widely used indicator is the Peak Signal-to-Noise Ratio (PSNR), which assesses the similarity between the original host signals and the embedded signals. The PSNR is defined for vectors as:
\begin{equation*}
	PSNR = 10 \times \log_{10}\left(\frac{{\text{MaxValue}}^2}{\text{MSE}}\right).
\end{equation*}
Here, \emph{MaxValue} represents the maximum value that the host signals can reach. In this simulation, since the host signals are the pixels of images, we set \emph{MaxValue} to be ${2^{8}} - 1$.

Another metric is the Percentage Residual Difference (PRD), which measures the rate of change between the original value and the difference, given by:
\begin{equation*}
	PRD = \sqrt{\frac{\sum_{i=1}^{M} | \mathbf{s}_k - \mathbf{s}_{w,k} |^2}{\sum_{i=1}^{M} | \mathbf{s}_k |^2}}.
\end{equation*}
Here, $\mathbf{s}_k$ denotes the original value and $\mathbf{s}{w,k}$ represents the corresponding difference.

    \begin{table*}[t!]
		\caption{The MSE, PSNR, and PRD performance of different algorithms when embedding messages to different parts of the DCT domain.}
		
		\begin{center}
			\scalebox{1}{
				\begin{tabular}{|c|c|ccccccccc|}
					\hline
					\multicolumn{2}{|c|} {} & \multicolumn{3}{c|}{$A_2$} & \multicolumn{3}{c|}{$D_4$}  & \multicolumn{3}{c|}{$E_8$}\\
					\hline
					\multicolumn{2}{|c|} {} & MSE & PSNR & \multicolumn{1}{c|}{PRD} & MSE & PSNR & \multicolumn{1}{c|}{PRD} & MSE & PSNR & PRD \\
					\hline
					\multirow{3}{*}{QIM} & low & 1.553 & 46.35 & 0.028 & 1.395 & 46.68 & 0.025 & 1.251 & 47.15 & 0.022\\
					\multirow{3}{*}{} & mid & 0.941 & 48.39 & 0.020 & 0.885 & 48.68 & 0.018 & 0.764 & 49.09 & 0.016\\
			        \multirow{3}{*}{} & high & 0.563 & 50.71 & 0.018 & 0.495 & 51.18 & 0.015 & 0.451 & 51.58 & 0.013\\
				    
					\hline
			    	\rowcolor{gray!40} \multirow{3}{*}{} & low & 1.074 & 47.82 & 0.020 & 0.868 & 48.74 & 0.019 & 0.727 & 49.51 & 0.016 \\
			   		\rowcolor{gray!40} \multirow{3}{*}{ } CA-QIM & mid & 0.706 & 49.39 & 0.016 & 0.511 & 51.04 & 0.014 & 0.454 & 51.56 & 0.013\\
			    	\rowcolor{gray!40} \multirow{3}{*}{ } & high & 0.415 & 51.96 & 0.014 & 0.327 & 52.89 & 0.011 & 0.231 & 54.41 & 0.011\\
			    	
					\hline
					\multirow{3}{*}{MD-QIM} & low & 0.272 & 53.79 & 0.014 & 0.219 & 54.62 & 0.012 & 0.174 & 55.72 & 0.011\\
					\multirow{3}{*}{ } & mid & 0.181 & 55.55 & 0.009 & 0.143 & 56.57 & 0.008 & 0.104 & 57.96 & 0.007\\
				    \multirow{3}{*}{ } & high & 0.094 & 58.39 & 0.007 & 0.075 & 59.38 & 0.004 & 0.063 & 60.13 & 0.004\\
				    
					\hline
					\rowcolor{gray!40} \multirow{3}{*}{} & low & 0.213 & 54.84 & 0.009 & 0.177 & 55.65 & 0.007 & 0.143 & 56.57 & 0.006\\
			   		\rowcolor{gray!40} \multirow{3}{*}{ }CAMD-QIM & mid & 0.138 & 56.73 & 0.007 & 0.098 & 58.21 & 0.006 & 0.076 & 59.32 & 0.005\\
			    	\rowcolor{gray!40} \multirow{3}{*}{ } & high & 0.047 & 61.41 & 0.006 & 0.035 & 62.69 & 0.005 & 0.024 & 64.32 & 0.003\\
					\hline
				\end{tabular}
			}
			\label{tab1}
		\end{center}
	\end{table*}
	
    \begin{figure*}[ht]
		\centering
		\subfloat[]{\includegraphics[width=0.13\linewidth]{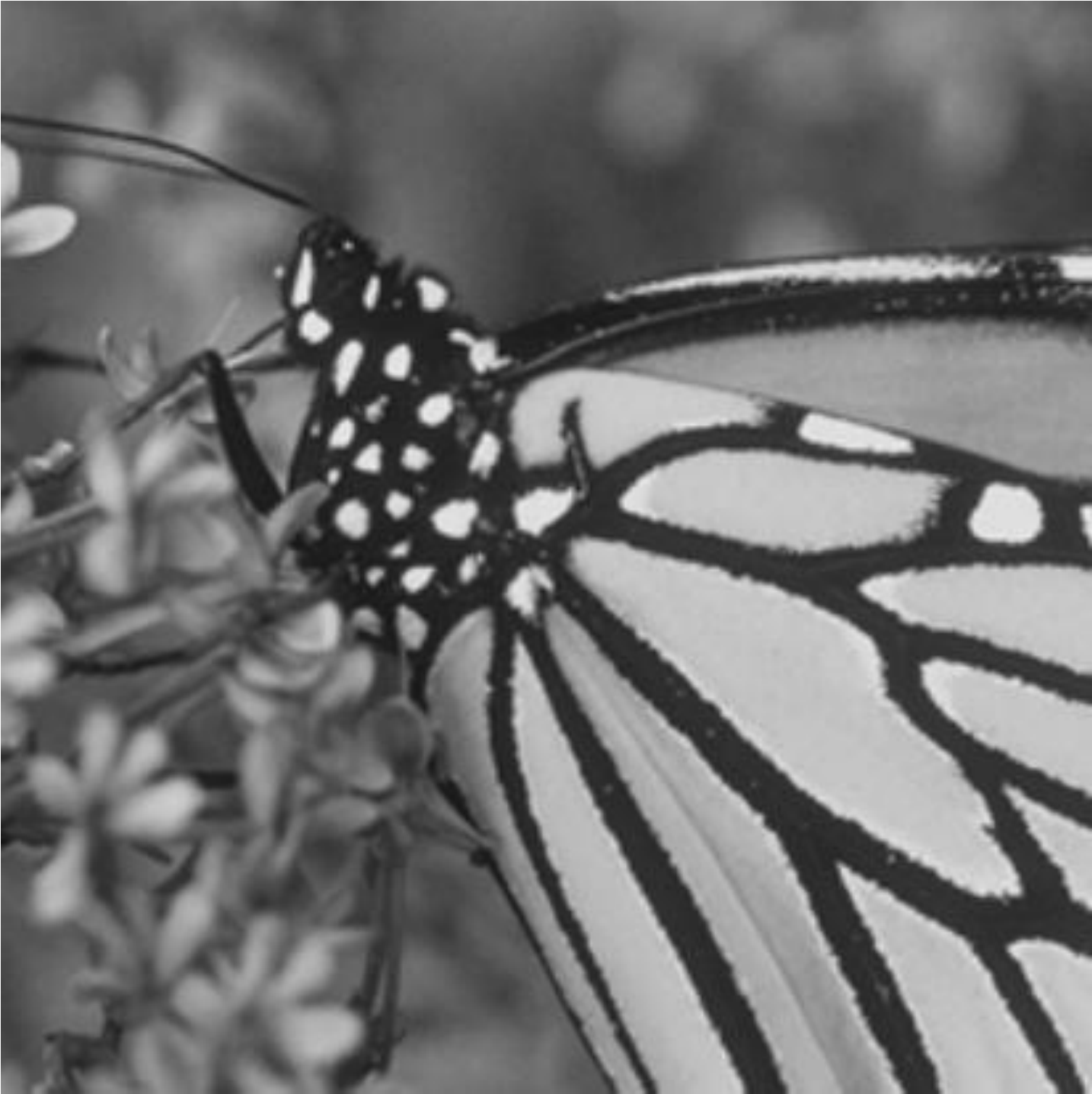}
		\includegraphics[width=0.13\linewidth]{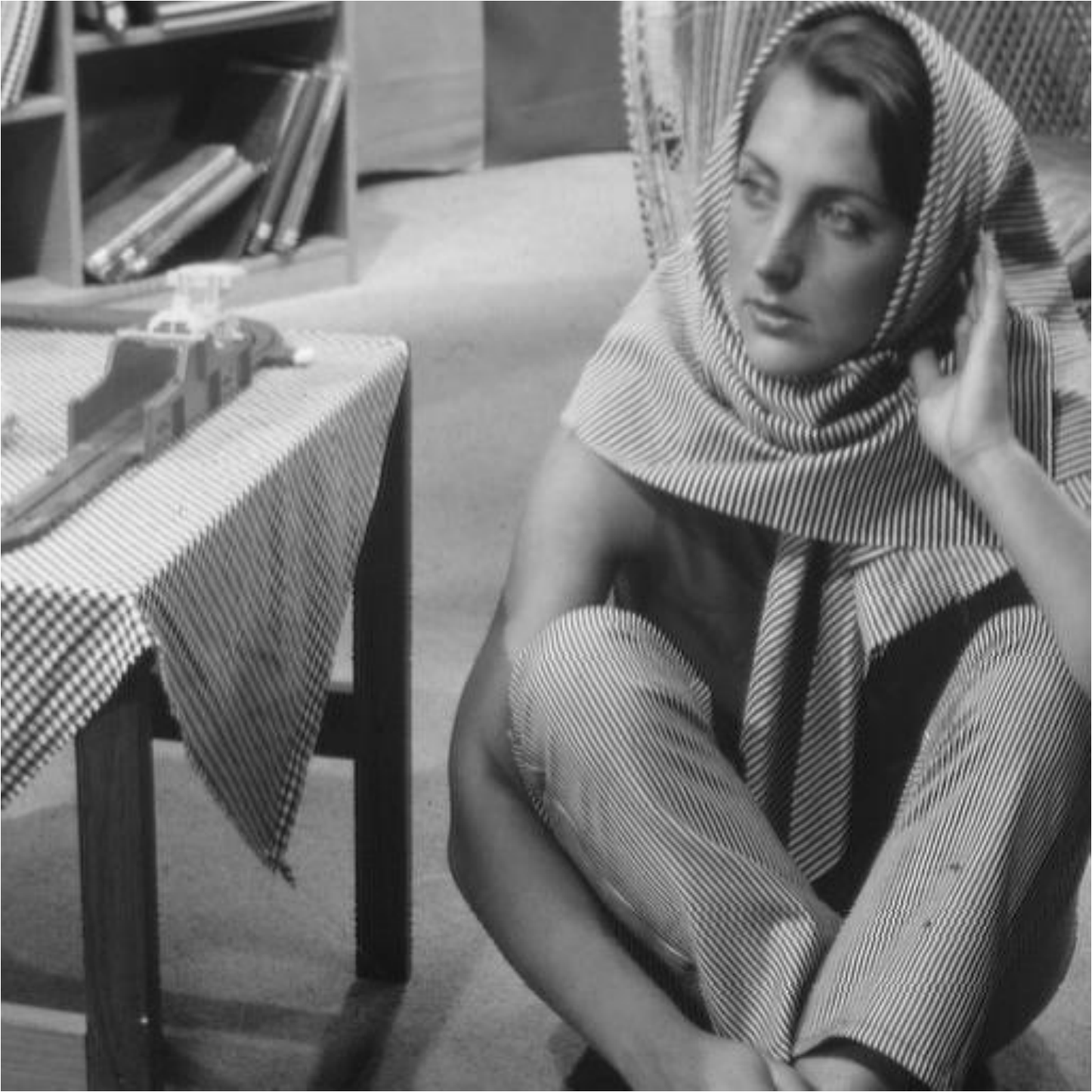}
		\includegraphics[width=0.13\linewidth]{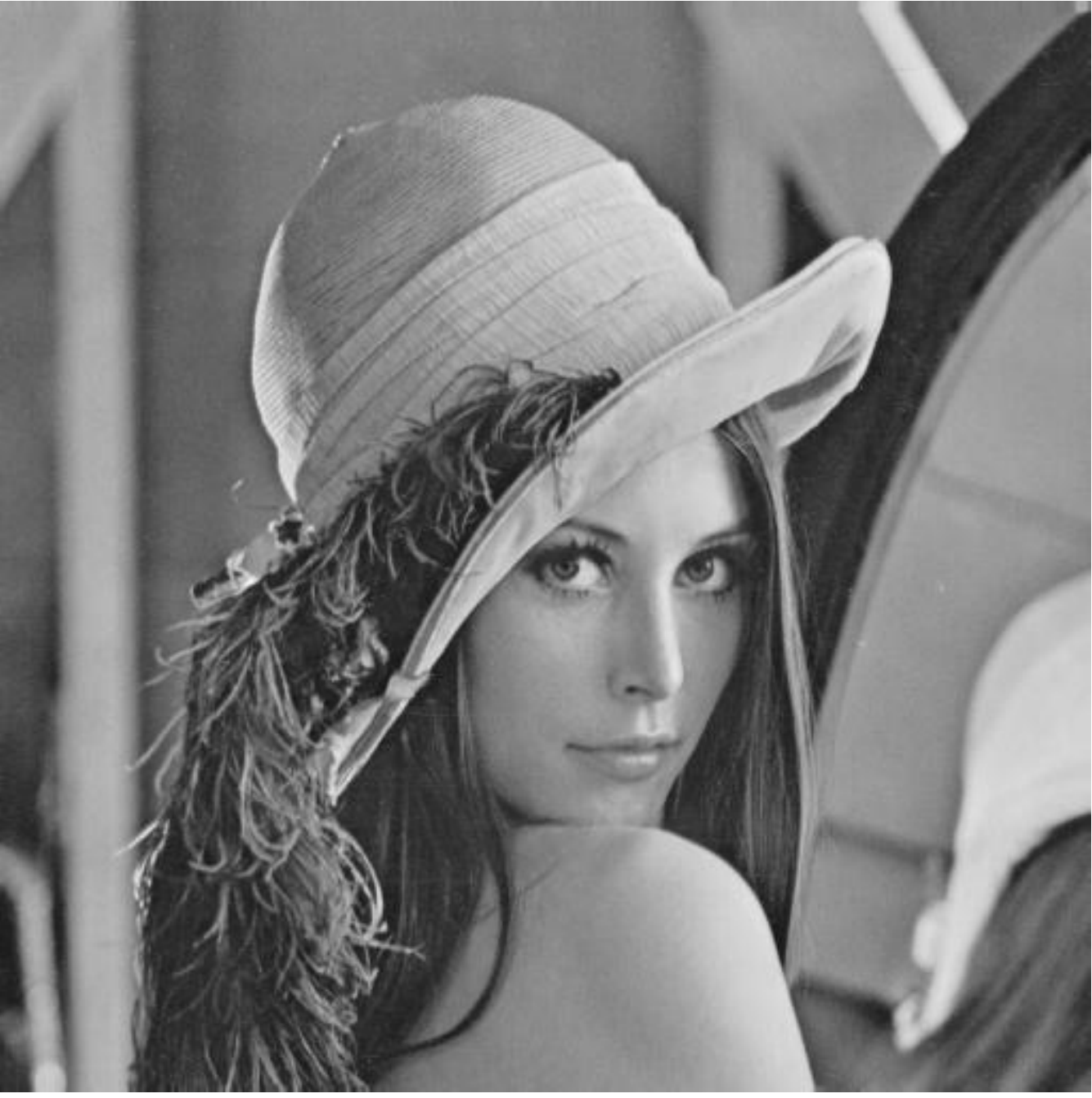}}
		\hspace{0.5cm}
		\subfloat[]{\includegraphics[width=0.13\linewidth]{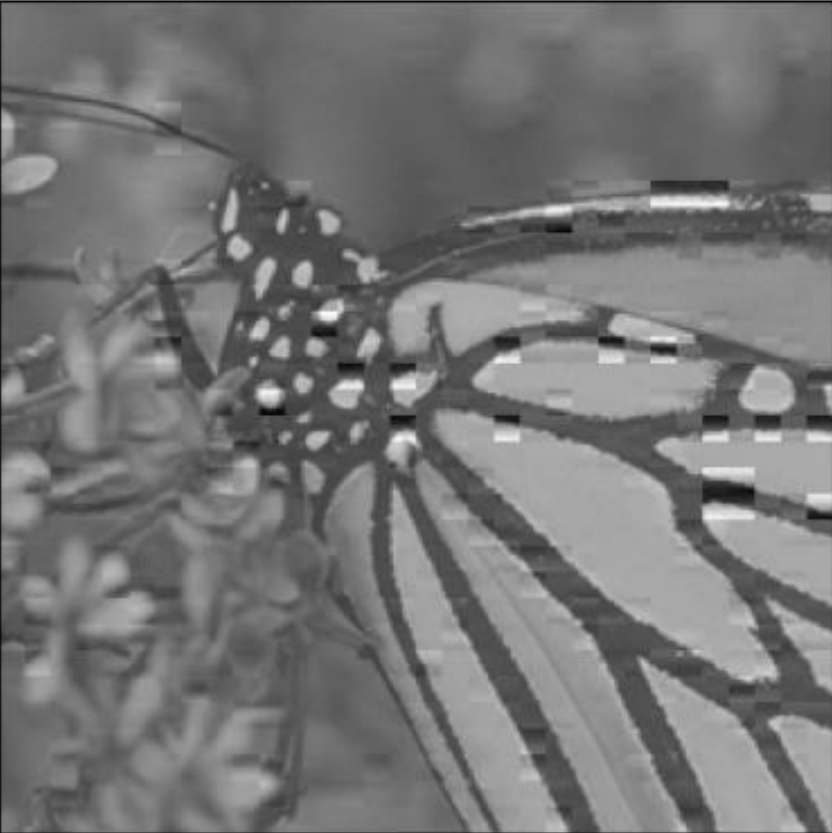}
		\includegraphics[width=0.13\linewidth]{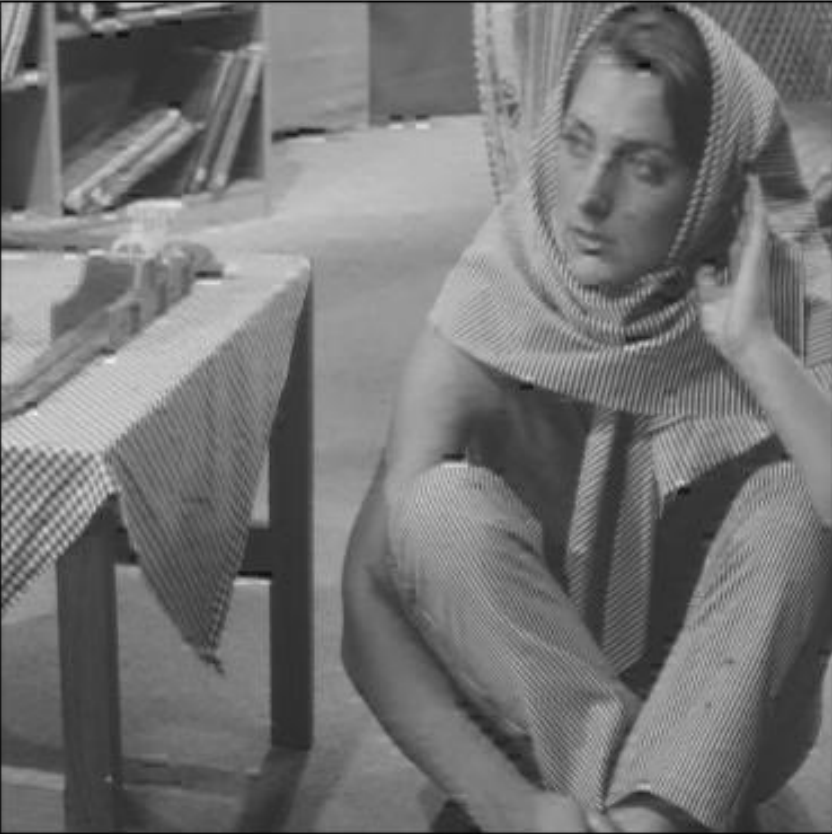}
		\includegraphics[width=0.13\linewidth]{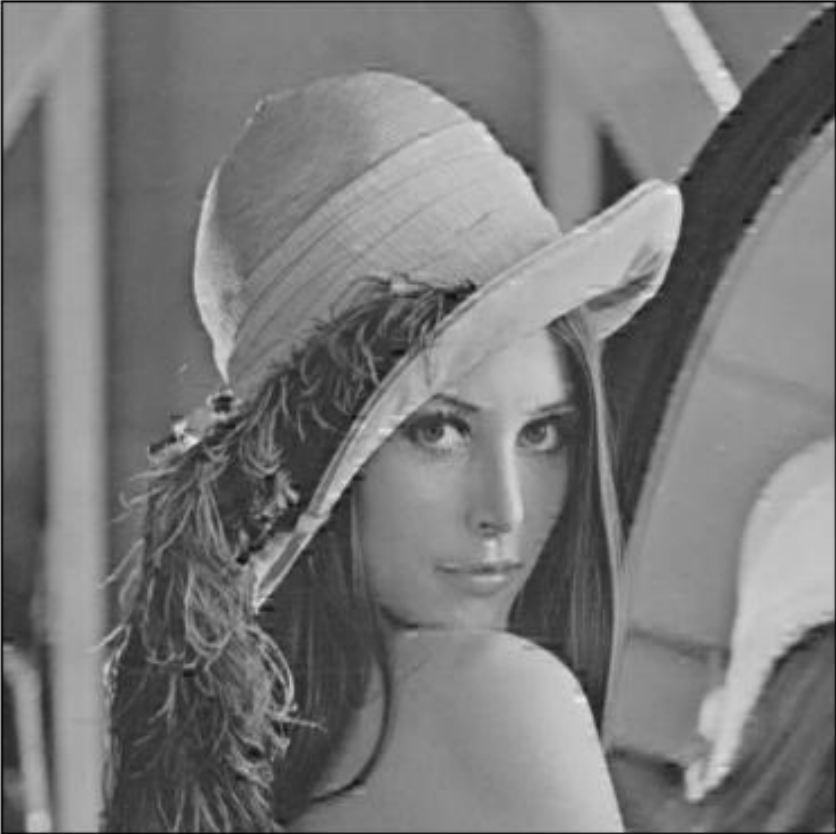}}
		\hspace{0.5cm}
		\subfloat[]{\includegraphics[width=0.13\linewidth]{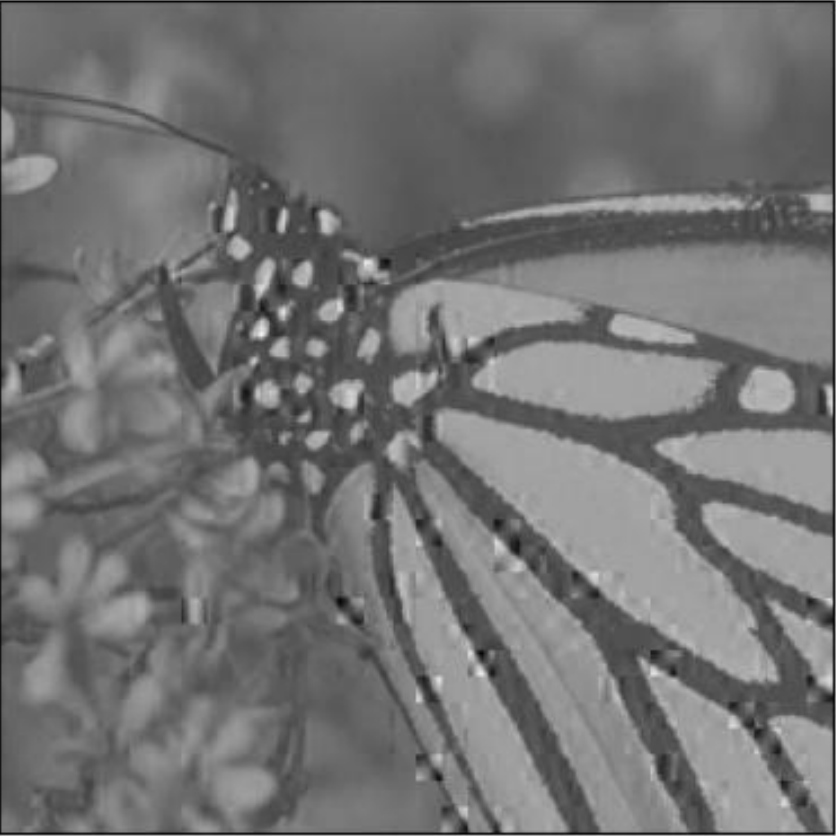}
		\includegraphics[width=0.13\linewidth]{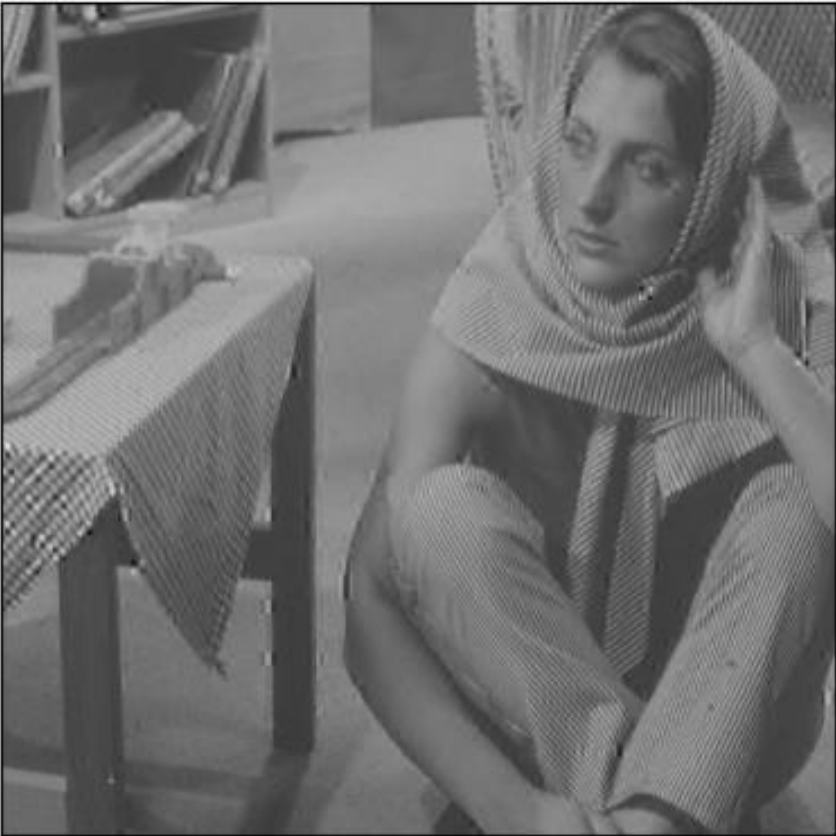}
		\includegraphics[width=0.13\linewidth]{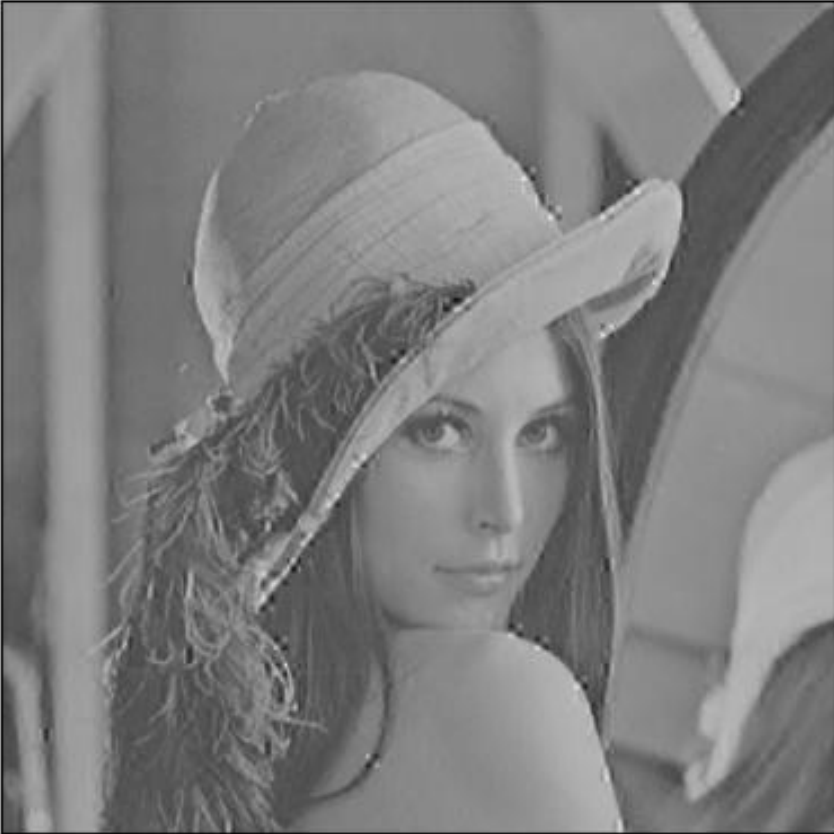}}
		\hspace{0.5cm}
		\subfloat[]{\includegraphics[width=0.13\linewidth]{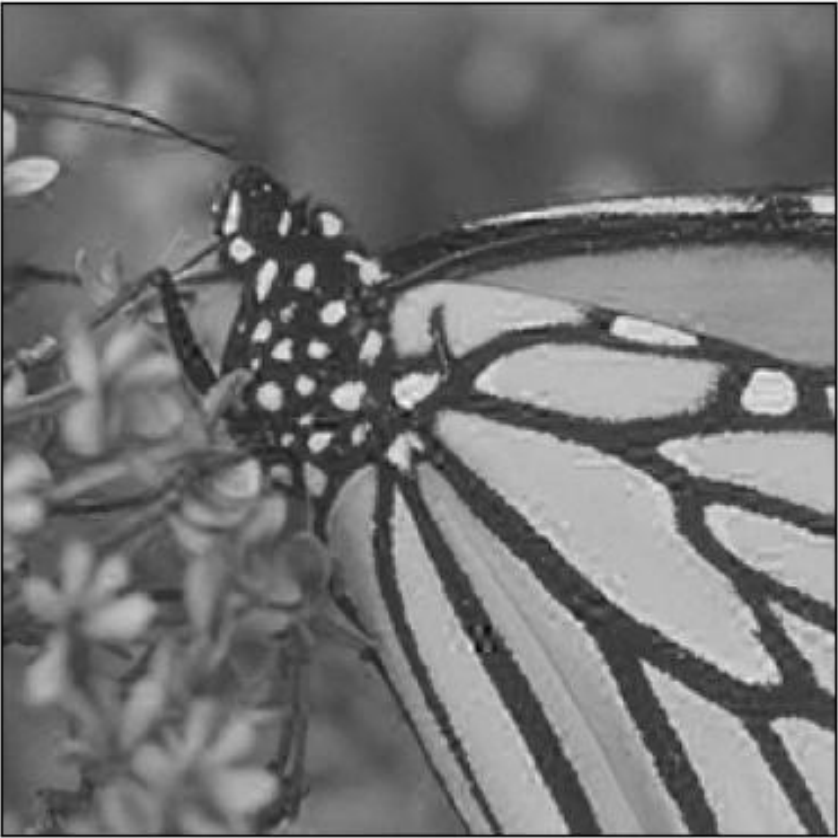}
		\includegraphics[width=0.13\linewidth]{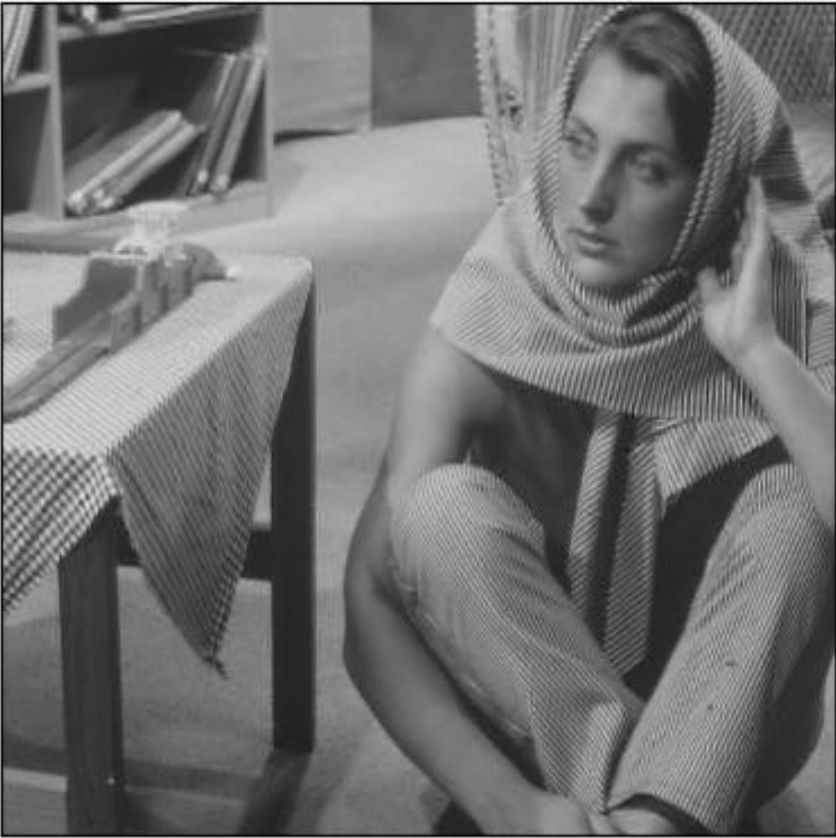}
		\includegraphics[width=0.13\linewidth]{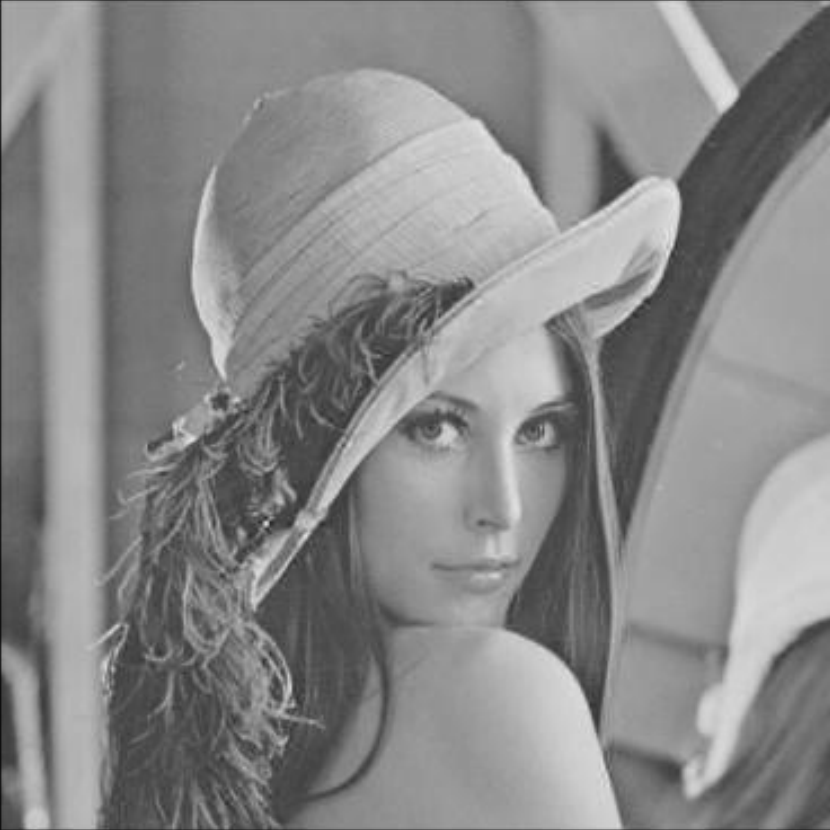}}
		\caption{Effects of embedding watermarks by using CA-QIM over different lattices. (a) Original image (b) $\Lambda_f=A_2$ (c)  $\Lambda_f=D_4$ (d)  $\Lambda_f=E_8$.}
		\label{spaital_fig1}
	\end{figure*}
	
    \begin{figure*}[ht]
		\centering
		\subfloat[]{\includegraphics[width=0.13\linewidth]{fig/fig6a.pdf}
		\includegraphics[width=0.13\linewidth]{fig/fig6e.pdf}
		\includegraphics[width=0.13\linewidth]{fig/fig6i.pdf}}
		\hspace{0.5cm}
		\subfloat[]{\includegraphics[width=0.13\linewidth]{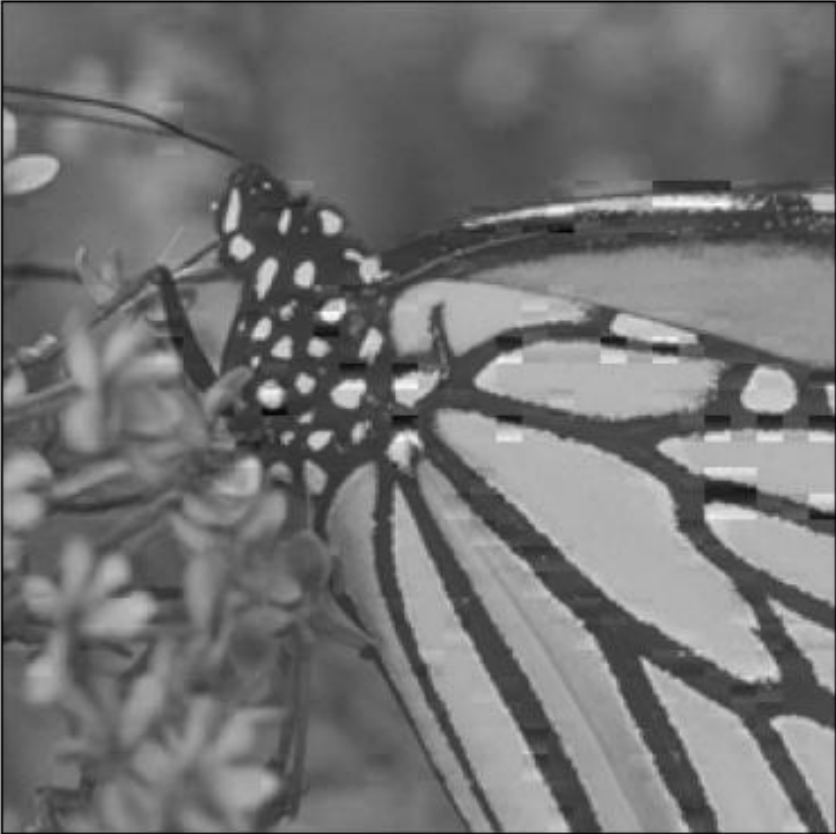}
		\includegraphics[width=0.13\linewidth]{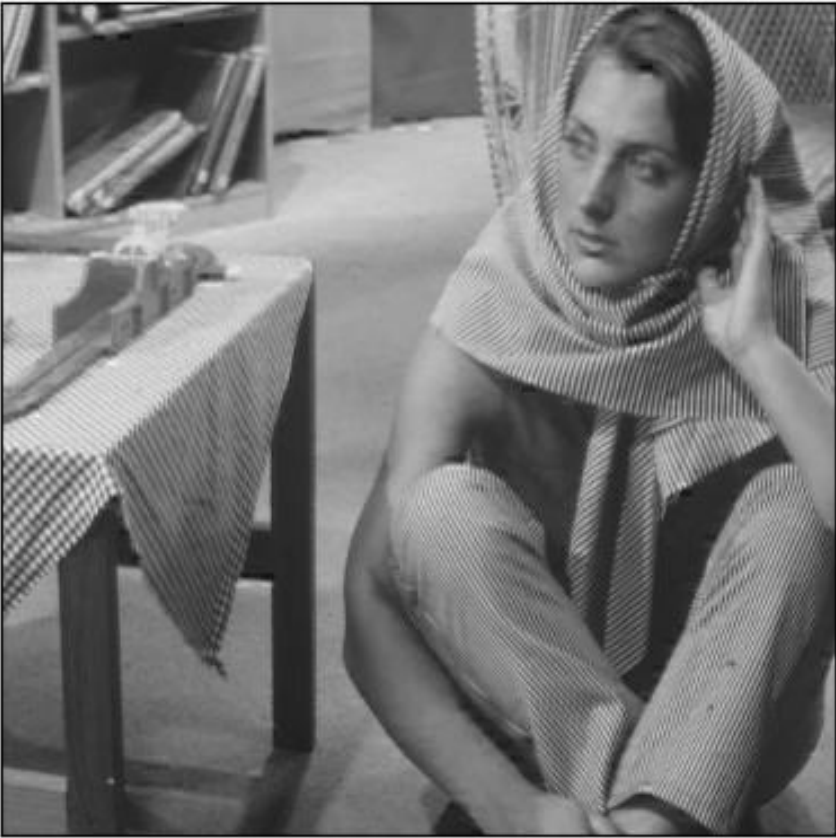}
		\includegraphics[width=0.13\linewidth]{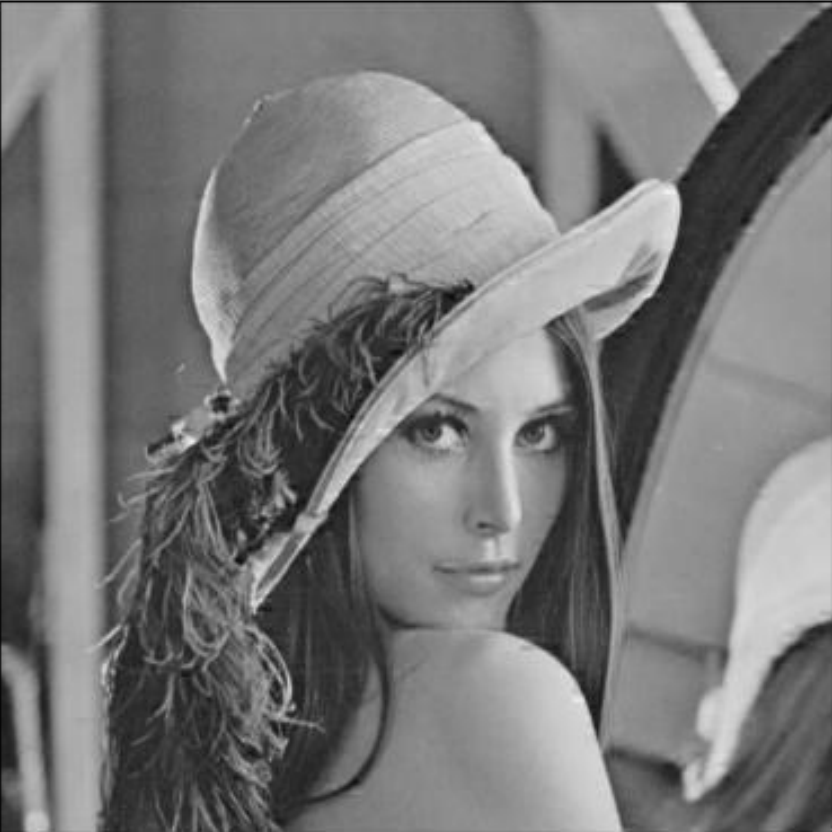}}
		\hspace{0.5cm}
		\subfloat[]{\includegraphics[width=0.13\linewidth]{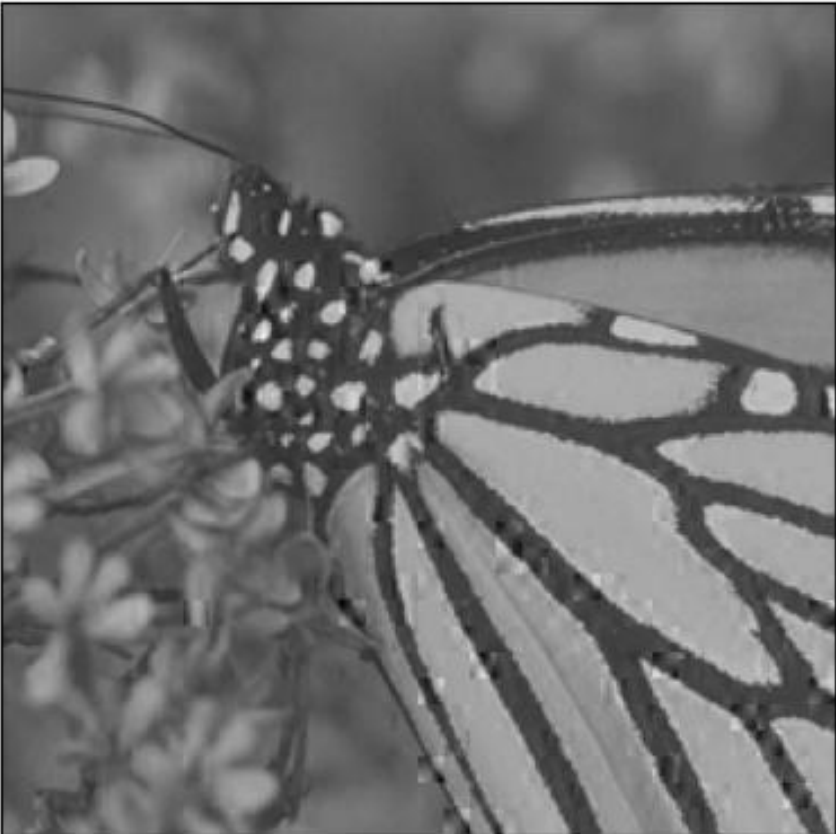}
		\includegraphics[width=0.13\linewidth]{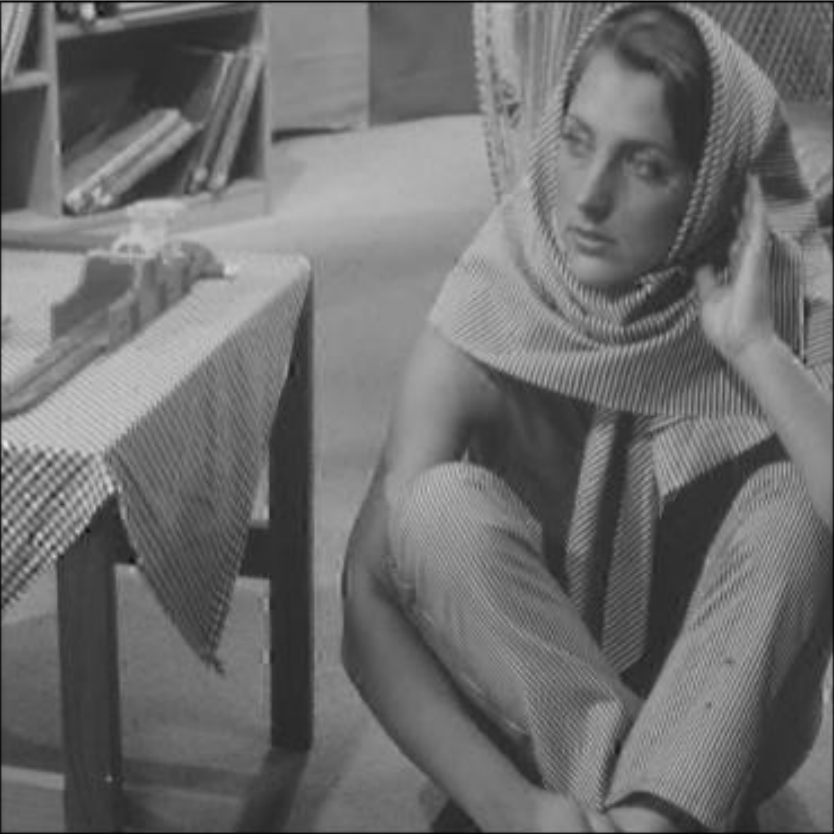}
		\includegraphics[width=0.13\linewidth]{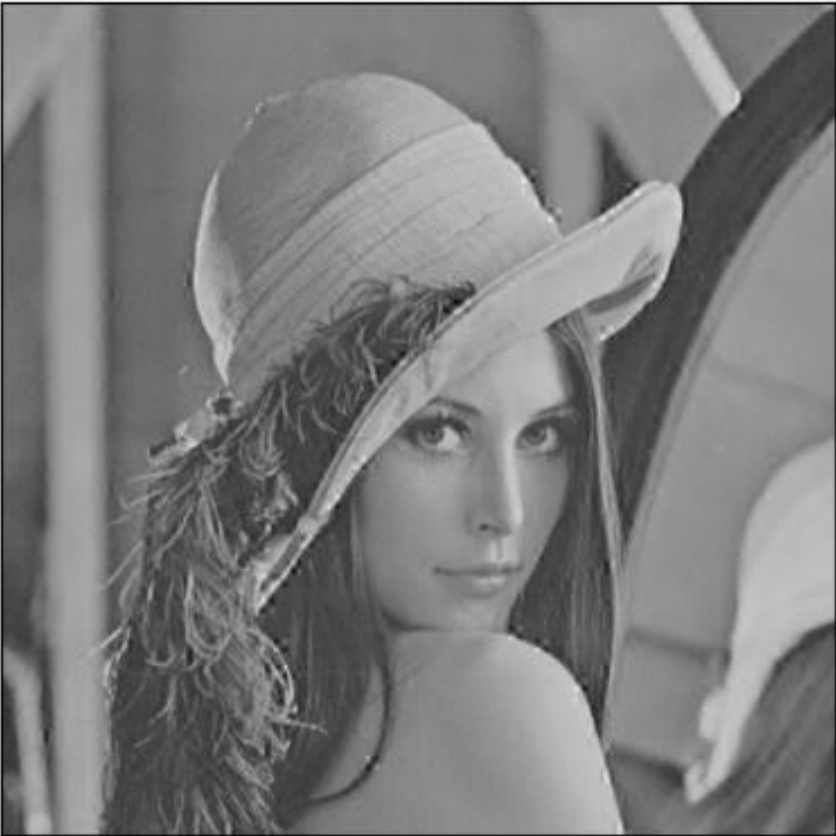}}
		\hspace{0.5cm}
		\subfloat[]{\includegraphics[width=0.13\linewidth]{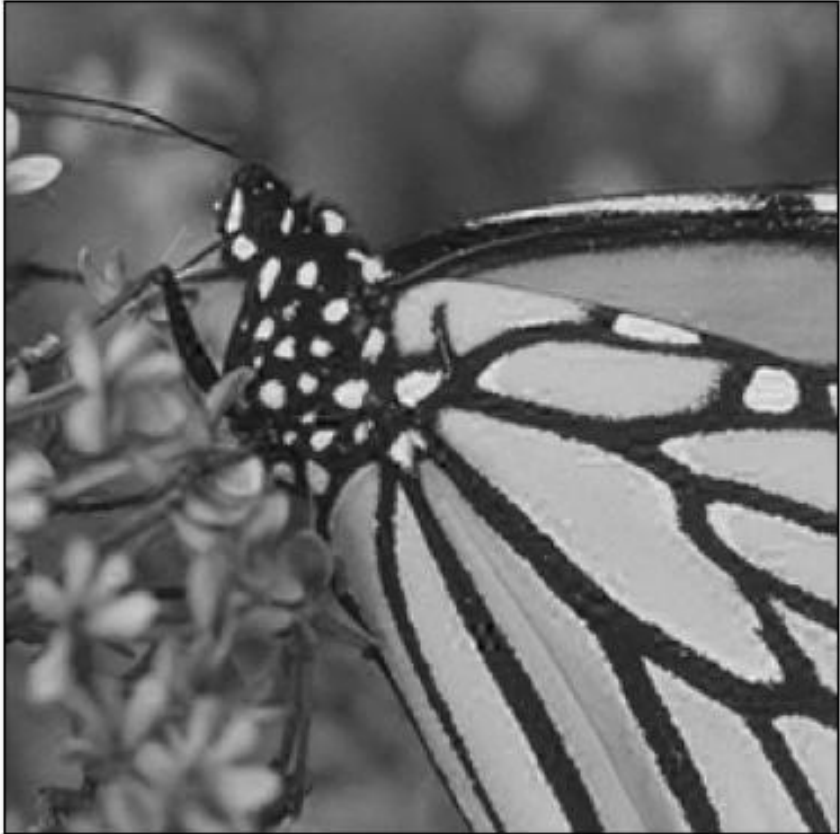}
		\includegraphics[width=0.13\linewidth]{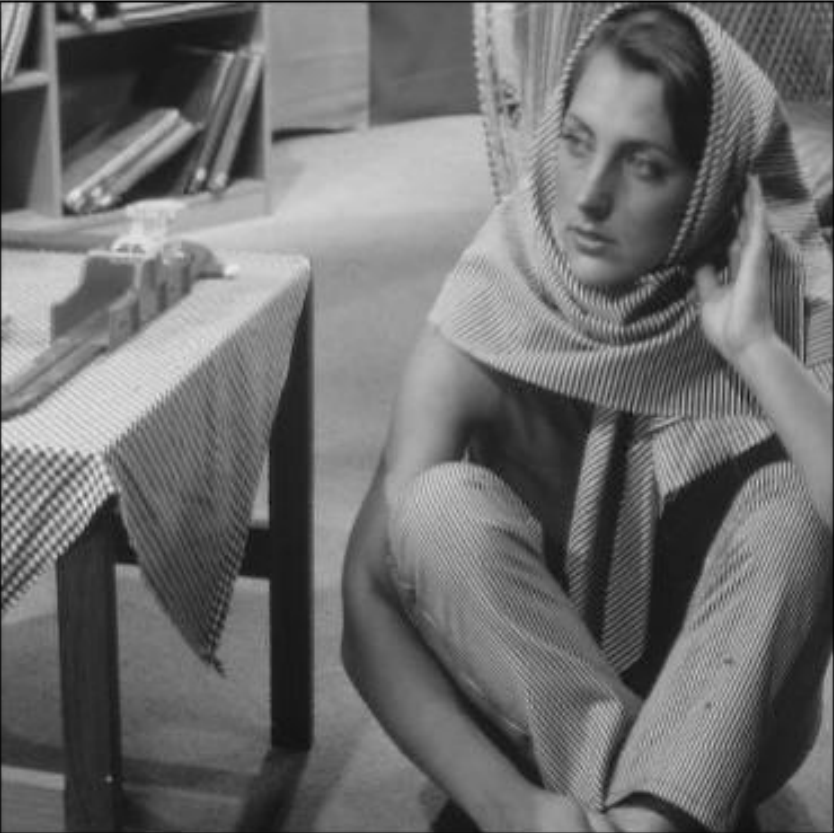}
		\includegraphics[width=0.13\linewidth]{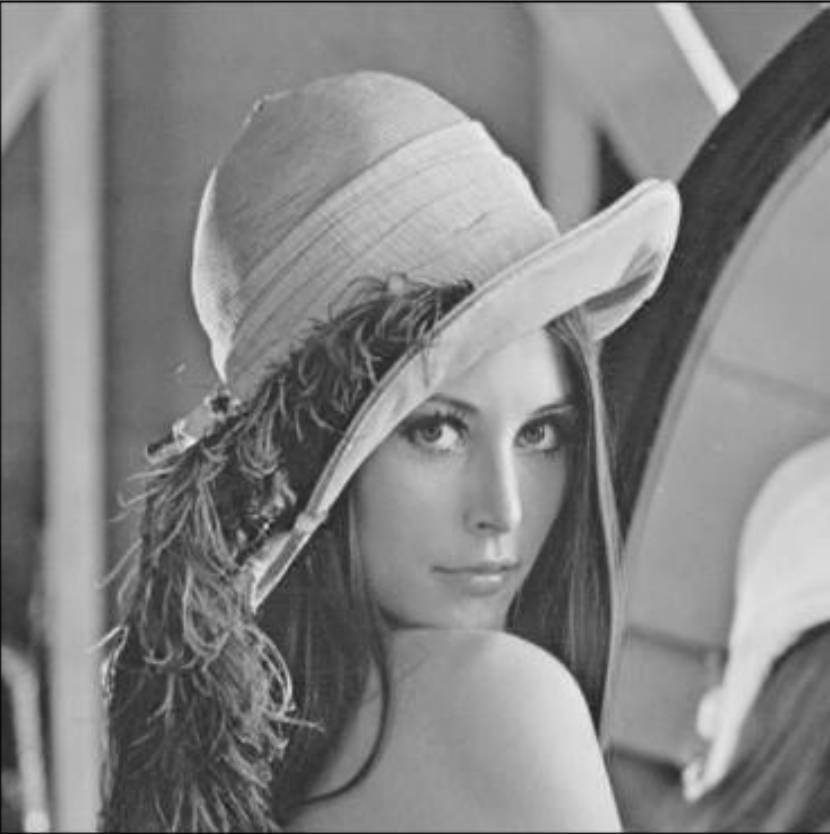}}
		\caption{Effects of embedding watermarks by using CAMD-QIM over different lattices. (a) Original image (b) $\Lambda_f=A_2$ (c)  $\Lambda_f=D_4$ (d)  $\Lambda_f=E_8$.}
		\label{spaital_fig2}
	\end{figure*}   
	
	In Table \ref{tab1}, we present the average values of MSE, PSNR, and PRD in the frequency domain after embedding using four benchmark algorithms: QIM, CA-QIM, MD-QIM, and CAMD-QIM on the images. Several conclusions can be drawn from the table:
	\begin{enumerate}
		\item 	Across all lattices, CA-QIM and CAMD-QIM consistently outperform QIM and MD-QIM in terms of the metrics. This suggests that CA-QIM exhibits lower distortion compared to the other algorithms.
		
	\item	Analyzing the metrics with different lattices reveals that the embedding process is influenced differently by the dimensions of the lattices.
		
	\item	CAMD-QIM employs the MD method to enhance imperceptibility, resulting in better performance than CA-QIM.
	\end{enumerate}

	Considering the impact of DCT and image pixels, the MSE increases when the image is converted from the frequency domain to the spatial domain. This increase is larger than the theoretical value mentioned in Section \ref{Distortion Analysis} (Distortion Analysis). 	Figures \ref{spaital_fig1} and \ref{spaital_fig2}  displays sample images by using different lattices for the embedding of CA-QIM and CAMD-QIM.

    \begin{table*}[t!]
		\caption{
		The MSE performance of different QIM variants when embedding messages to different frequencies of the DCT domain.
	}
		
		\begin{center}
			\scalebox{1.2}{
				\begin{tabular}{|c|c|ccccccccc|}
					\hline
					\multicolumn{2}{|c|} {} & \multicolumn{3}{c|}{$A_2$} & \multicolumn{3}{c|}{$D_4$}  & \multicolumn{3}{c|}{$E_8$}\\
					\hline
					\multicolumn{2}{|c|} {payload (bits/dim)} & 1 & 2 & \multicolumn{1}{c|}{3} & 1 & 2 & \multicolumn{1}{c|}{3} & 1 & 2 & 3 \\
					\hline
					\multirow{3}{*}{QIM} & low & 1.5061 & 1.5408 & 1.5932 & 1.4979 & 1.5312 & 1.5782 & 1.4978 & 1.5309 & 1.5770\\
					\multirow{3}{*}{ } & mid & 0.5061 & 0.5468 & 0.5883 & 0.5039 & 0.5419 & 0.5816 & 0.4907 & 0.5258 & 0.5606\\
			        \multirow{3}{*}{ } & high & 0.1553 & 0.2012 & 0.2704 & 0.1441 & 0.2164 & 0.2416 & 0.1475 & 0.1872 & 0.2371\\
			        
					\hline
			    	\rowcolor{gray!40} \multirow{3}{*}{} & low & 1.0061 & 1.1304 & 1.1821 & 1.0139 & 1.0634 & 1.0966 & 0.8378 & 0.9037 & 1.0674\\
			   		\rowcolor{gray!40} \multirow{3}{*}{ }CA-QIM & mid & 0.5899 & 0.5961 & 0.6601 & 0.5186 & 0.5411 & 0.5655 & 0.4201 & 0.4764 & 0.5004\\
			    	\rowcolor{gray!40} \multirow{3}{*}{ } & high & 0.1182 & 0.1752 & 0.2288 & 0.1012 & 0.1336 & 0.1532 & 0.1294 & 0.1480 & 0.1936\\
			    	
					\hline
					\multirow{3}{*}{MD-QIM} & low & 0.8771 & 0.8925 & 0.9311 & 0.8693 & 0.8970 & 0.8431 & 0.8568 & 0.8785 & 0.8917\\
					\multirow{3}{*}{ } & mid & 0.3795 & 0.4857 & 0.4831 & 0.3590 & 0.3747 & 0.4365 & 0.3978 & 0.3840 & 0.4101\\
				    \multirow{3}{*}{ } & high & 0.0197 & 0.0674 & 0.1083 & 0.1617 & 0.3721 & 0.4170 & 0.3043 & 0.4672 & 0.4996\\
				    
					\hline
			    	\rowcolor{gray!40} \multirow{3}{*}{} & low & 0.7371 & 0.7607 & 0.7954 & 0.7193 & 0.7548 & 0.7892 & 0.7068 & 0.7313 & 0.7602\\
			   		\rowcolor{gray!40} \multirow{3}{*}{ }CAMD-QIM & mid & 0.4486 & 0.4689 & 0.4703 & 0.4590 & 0.4530 & 0.5194 & 0.4782 & 0.4827 & 0.4870\\
			    	\rowcolor{gray!40} \multirow{3}{*}{ } & high & 0.0184 & 0.0585 & 0.1042 & 0.1498 & 0.2193 & 0.3049 & 0.1963 & 0.2389 & 0.2999\\
					\hline
				\end{tabular}
			}
			\label{tab3}
		\end{center}
	\end{table*}

    \begin{table*}[t!]
		\caption{The PSNR and SSIM performance of different QIM variants when embedding messages to different frequencies of the DCT domain.}
		\begin{center}
			\scalebox{1.3}{
				\begin{tabular}{|c|c|ccccccccc|}
					\hline
					\multicolumn{2}{|c|} {} & \multicolumn{3}{c|}{$A_2$} & \multicolumn{3}{c|}{$D_4$}  & \multicolumn{3}{c|}{$E_8$}\\
					
					\hline
					\multicolumn{2}{|c|} {payload (bits/dim)} & 1 & 2 & \multicolumn{1}{c|}{3} & 1 & 2 & \multicolumn{1}{c|}{3} & 1 & 2 & 3 \\
					\hline

					\multirow{2}{*}{low} & PSNR & 
				    51.0876 & 51.3114 & 51.3600 & 51.4156 & 51.0141 & 51.2356 & 51.1299 & 51.1291 & 51.0528  \\
					\multirow{2}{*}{ } & SSIM & 0.9585 & 0.9418 & 0.9346 & 0.9600 & 0.9368 & 0.9263 & 0.9716 & 0.9616 & 0.9554\\
					\hline

					\multirow{2}{*}{mid} & PSNR & 52.7065 & 51.4196 & 51.4069 & 51.5126 & 51.5695 & 50.9757 & 51.3341 & 51.2939 & 51.2552  \\
					\multirow{2}{*}{ } & SSIM & 0.9989 & 0.9975 & 0.9971 & 0.9963 & 0.9925 & 0.9908 & 0.9960 & 0.9899 & 0.9874 \\
					
					\hline
					\multirow{2}{*}{high} & PSNR & 65.4625 & 60.4534 & 57.9480 & 56.3737 & 52.4230 & 52.0562 & 53.4127 & 51.7070 & 51.1443  \\
					\multirow{2}{*}{ } & SSIM & 0.9999 & 0.9999 & 0.9998 & 0.9997 & 0.9992 & 0.9985 & 0.9994 & 0.9981 & 0.9956 \\
					\hline
				\end{tabular}
			}
			\label{tab4}
		\end{center}
	\end{table*}
	
	\subsection{Embedding Capacity Measurement}
	The embedding capacity is an important factor for evaluating the quality of the proposed method. In this subsection, we measure the embedding capacity by varying the number of messages and the positions of embedding AC coefficients. Since each block has 63 AC coefficients, the number of embeddable messages per block ranges from 1 to $\lfloor63/N\rfloor$. To comply with the payload limits, we embed one, two, and three labeling messages $\mathbf{m}$ into a block, respectively.
	
	Table 3 presents a more comprehensive set of metrics, including MSE in various scenarios. From Table 3, it can be observed that CAMD-QIM outperforms MD-QIM in low-dimensional lattices, and the distinction becomes more pronounced as the lattice dimension increases. Additionally, CAMD-QIM also performs better than MD-QIM at the same frequency.
	
	In addition to the three aforementioned metrics, Structural Similarity Index (SSIM) is introduced to provide a comprehensive evaluation. SSIM measures the similarity between two images and is calculated as follows:
	\begin{equation}
		SSIM(\mathbf{x},\mathbf{y}) = \frac{(2\mu_\mathbf{x}\mu_\mathbf{y}+c_1)(2\sigma_{\mathbf{xy}}+c_2)}{({\mu_\mathbf{x}}^2+{\mu_\mathbf{y}}^2+c_1)({\sigma_\mathbf{x}}^2+{\sigma_\mathbf{y}}^2+c_2)},
	\end{equation}
	where $\mu_\mathbf{x}$ and $\mu_\mathbf{y}$ represent the mean of images $\mathbf{x}$ and $\mathbf{y}$, respectively, and $\sigma_\mathbf{x}$ and $\sigma_\mathbf{y}$ represent the variances of images $\mathbf{x}$ and $\mathbf{y}$, respectively. The covariance of images $\mathbf{x}$ and $\mathbf{y}$ is denoted as $\sigma_\mathbf{xy}$.
	
	To visualize the impact of embedded messages on the image, we obtain new images after embedding the messages. Taking CA-QIM as an example, Table 4 presents the average PSNR and SSIM values obtained by comparing the new images with the original image in the spatial domain.
	
	From Table \ref{tab3}, it can be observed that the position of embedding has a significant impact on the images, especially in low frequencies. On the contrary, the effects of lattice dimension and the number of embedded messages are relatively small. However, it is still noticeable that as the lattice dimension increases, there is an increase in deviation, which may be caused by the inverse discrete cosine transform.
	
\begin{figure*}[ht]
		\centering
		\subfloat[]{\includegraphics[width=0.18\linewidth]{fig/fig6a.pdf}}
		\hspace{0.5cm}
		\subfloat[]{\includegraphics[width=0.18\linewidth]{fig/fig6b.pdf}}
		\hspace{0.5cm}
		\subfloat[]{\includegraphics[width=0.18\linewidth]{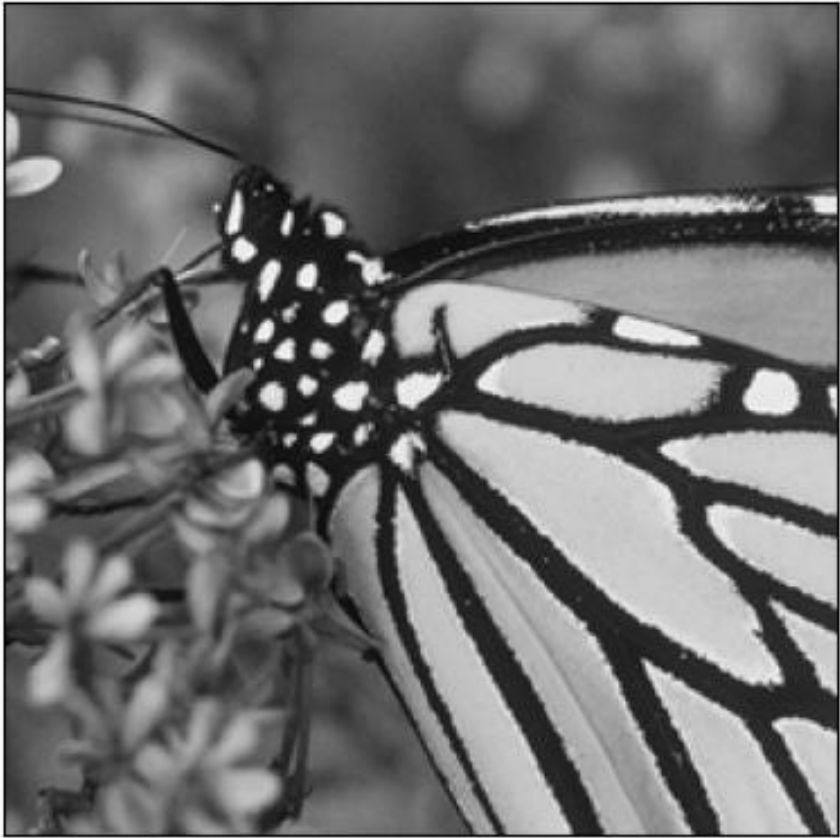}}
		\hspace{0.5cm}
		\subfloat[]{\includegraphics[width=0.18\linewidth]{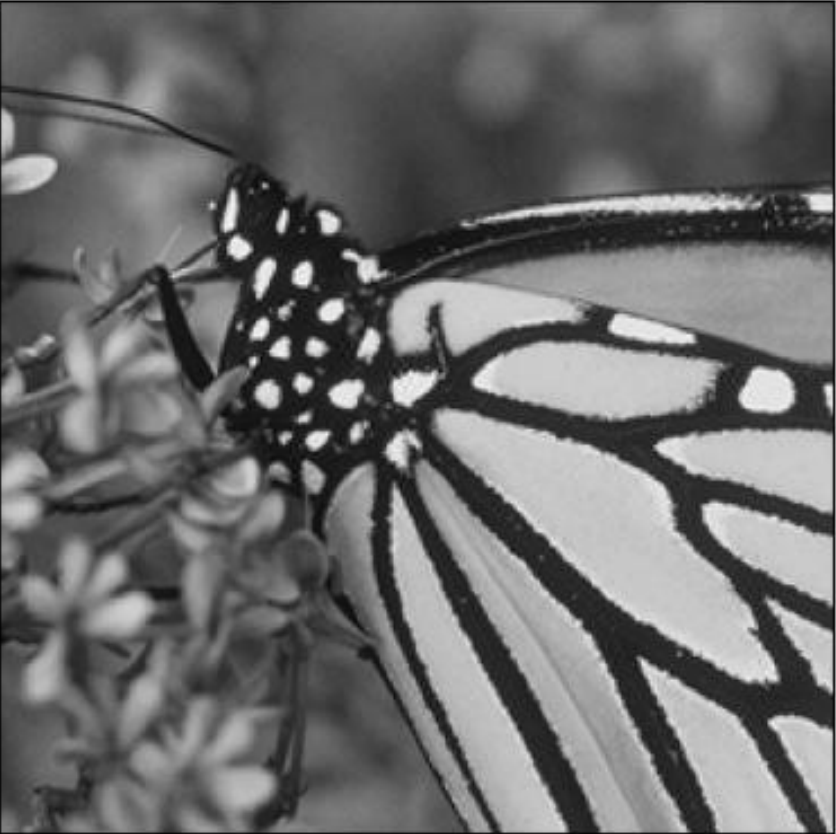}}
		\caption{Embedding in images via CA-QIM with $\Lambda_f=A_2$. (a) Original image (b) One message embedded in low frequency domain via $A_2$ (c) Two message embedded in medium frequency domain via $A_2$ (d)Three message embedded in high frequency domain via $A_2$.}
		\label{spaital_fig3}
	\end{figure*}

\begin{figure*}[ht]
		\centering
		\subfloat[]{\includegraphics[width=0.18\linewidth]{fig/fig6a.pdf}}
		\hspace{0.5cm}
		\subfloat[]{\includegraphics[width=0.18\linewidth]{fig/fig6b.pdf}}
		\hspace{0.5cm}
		\subfloat[]{\includegraphics[width=0.18\linewidth]{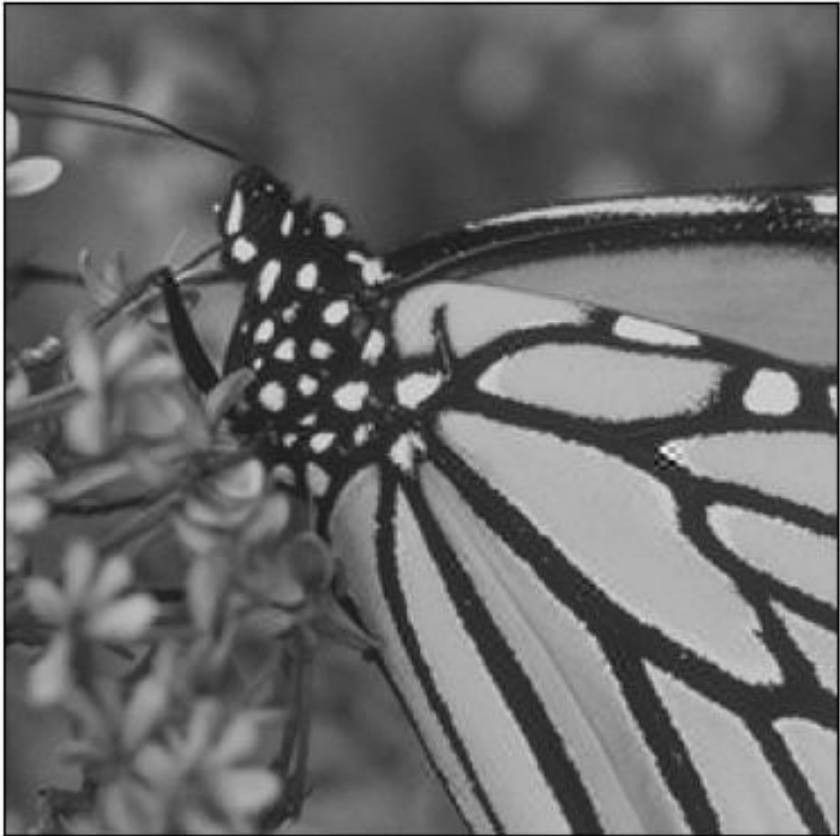}}
		\hspace{0.5cm}
		\subfloat[]{\includegraphics[width=0.18\linewidth]{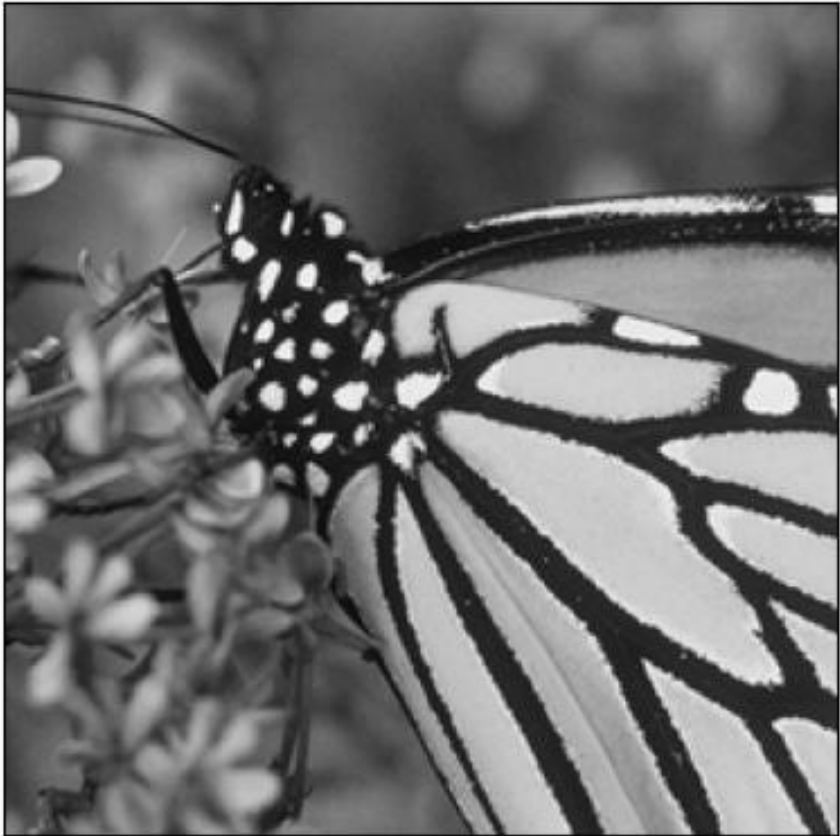}}
		\caption{Embedding in images via CA-QIM with $\Lambda_f=D_4$. (a) Original image (b) One message embedded in low frequency domain via $D_4$ (c) Two message embedded in medium frequency domain via $D_4$ 
		(d)Three message embedded in high frequency domain via $D_4$.}
		\label{spaital_fig4}
	\end{figure*}

\begin{figure*}[t]
		\centering
		\subfloat[]{\includegraphics[width=0.18\linewidth]{fig/fig6a.pdf}}
		\hspace{0.5cm}
		\subfloat[]{\includegraphics[width=0.18\linewidth]{fig/fig6d.pdf}}
		\hspace{0.5cm}
		\subfloat[]{\includegraphics[width=0.18\linewidth]{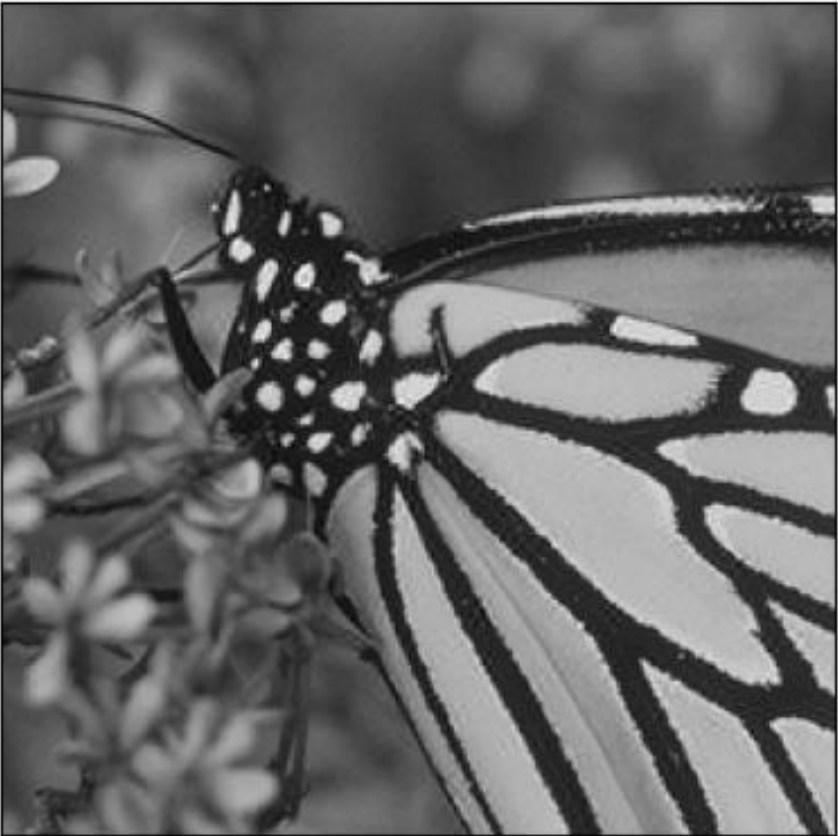}}
		\hspace{0.5cm}
		\subfloat[]{\includegraphics[width=0.18\linewidth]{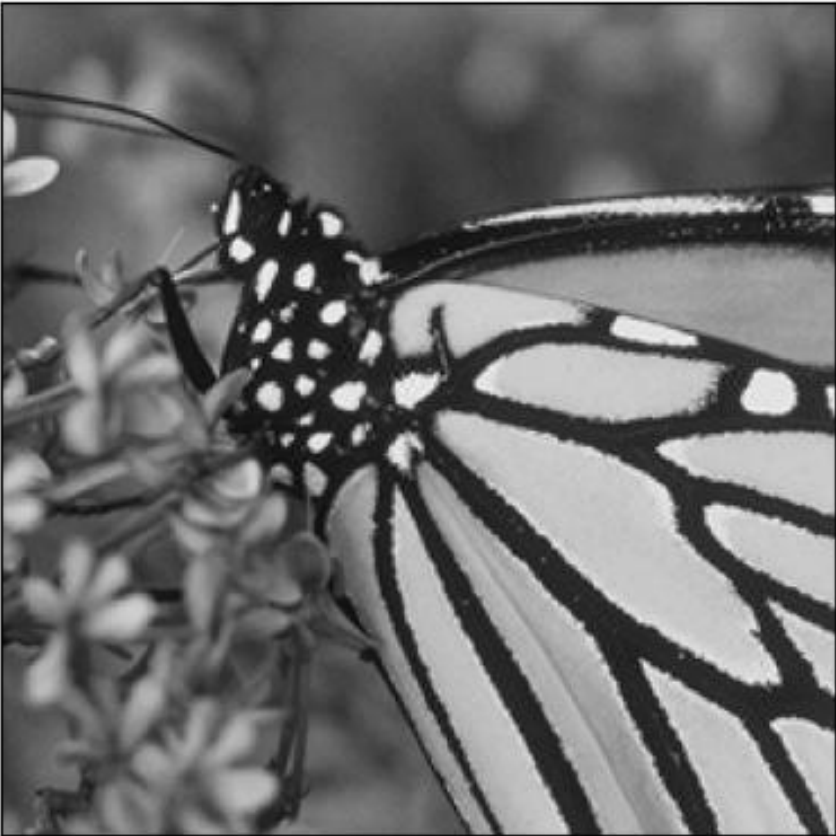}}
		\caption{Embedding in images via CA-QIM with $\Lambda_f=E_8$. (a) Original image (b) One message embedded in low frequency domain via $E_8$ (c) Two message embedded in medium frequency domain via $E_8$ 
		(d)Three message embedded in high frequency domain via $E_8$.}
	\label{spaital_fig5}	
	\end{figure*}

	Furthermore, from Figures \ref{spaital_fig3}, \ref{spaital_fig4}, and \ref{spaital_fig5}, it can be observed that embedding one message at the low frequency DCT domain  via $A_2$ already results in a noticeable difference in the new image compared to the original image. However, after embedding three messages in the medium and high frequencies in $A_2$, there is still no significant difference observed in the new image. Similarly, when embedding one message in low frequencies of $D_4$ and $E_8$, the difference from the original image is clearly visible. In the case of medium frequencies in $D_4$ and $E_8$, the new image shows a difference from the original image after embedding two messages, while three messages embedded in high frequencies are inconspicuous in both $D_4$ and $E_8$.
	
	In summary, due to the varying impact at different embedding sites, embedding in high or medium frequencies results in less distortion. Additionally, the embedding capacity is influenced by the embedding position. Higher frequencies allow for more messages to be embedded without detection. Therefore, the embedding position in the image should be selected as medium or high frequency.

\subsection{Symbol Error Rate Measurement}
\begin{figure*}[t]
		\centering
		\subfloat[]{\includegraphics[width=0.2\linewidth]{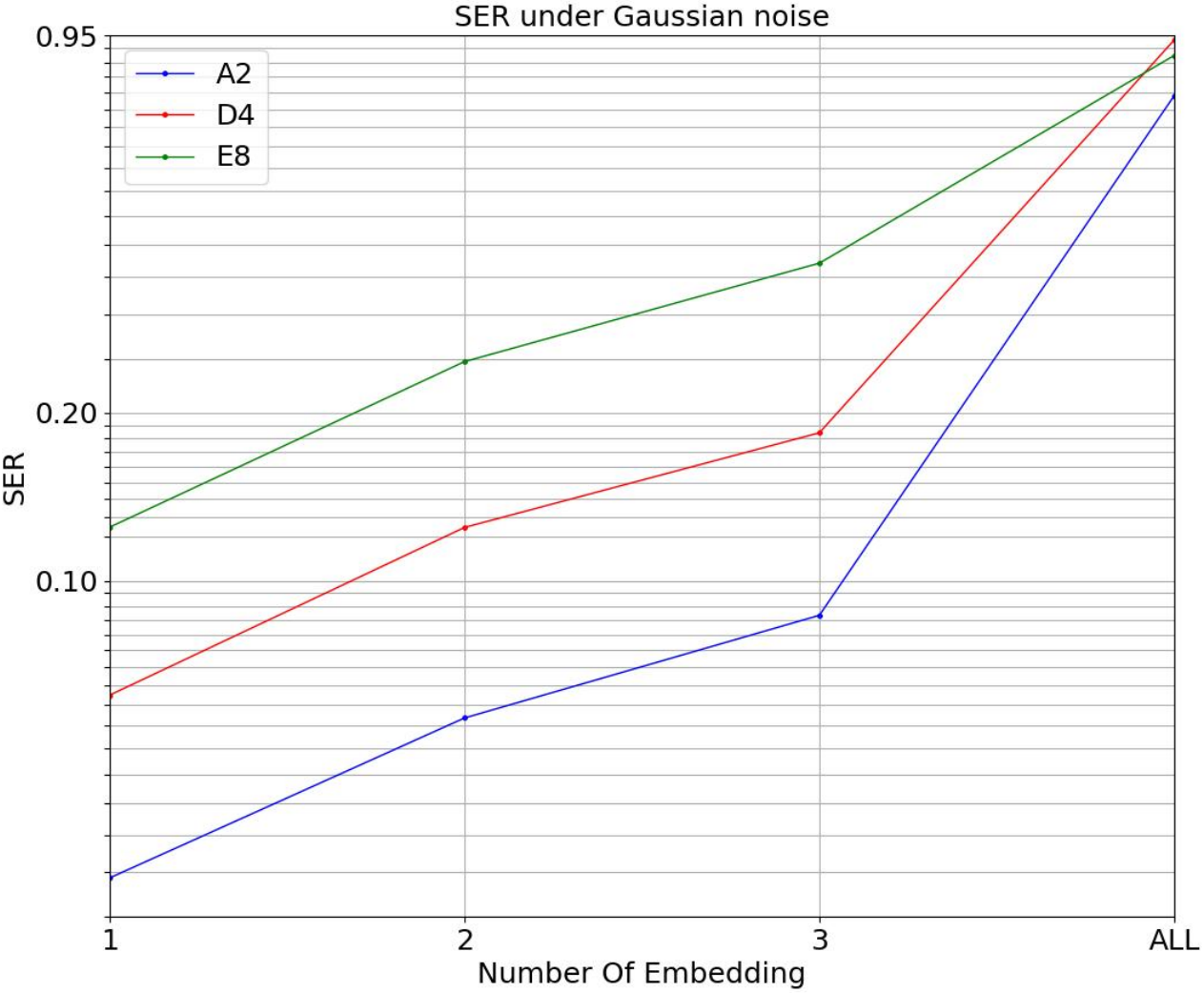}}
		\subfloat[]{\includegraphics[width=0.2\linewidth]{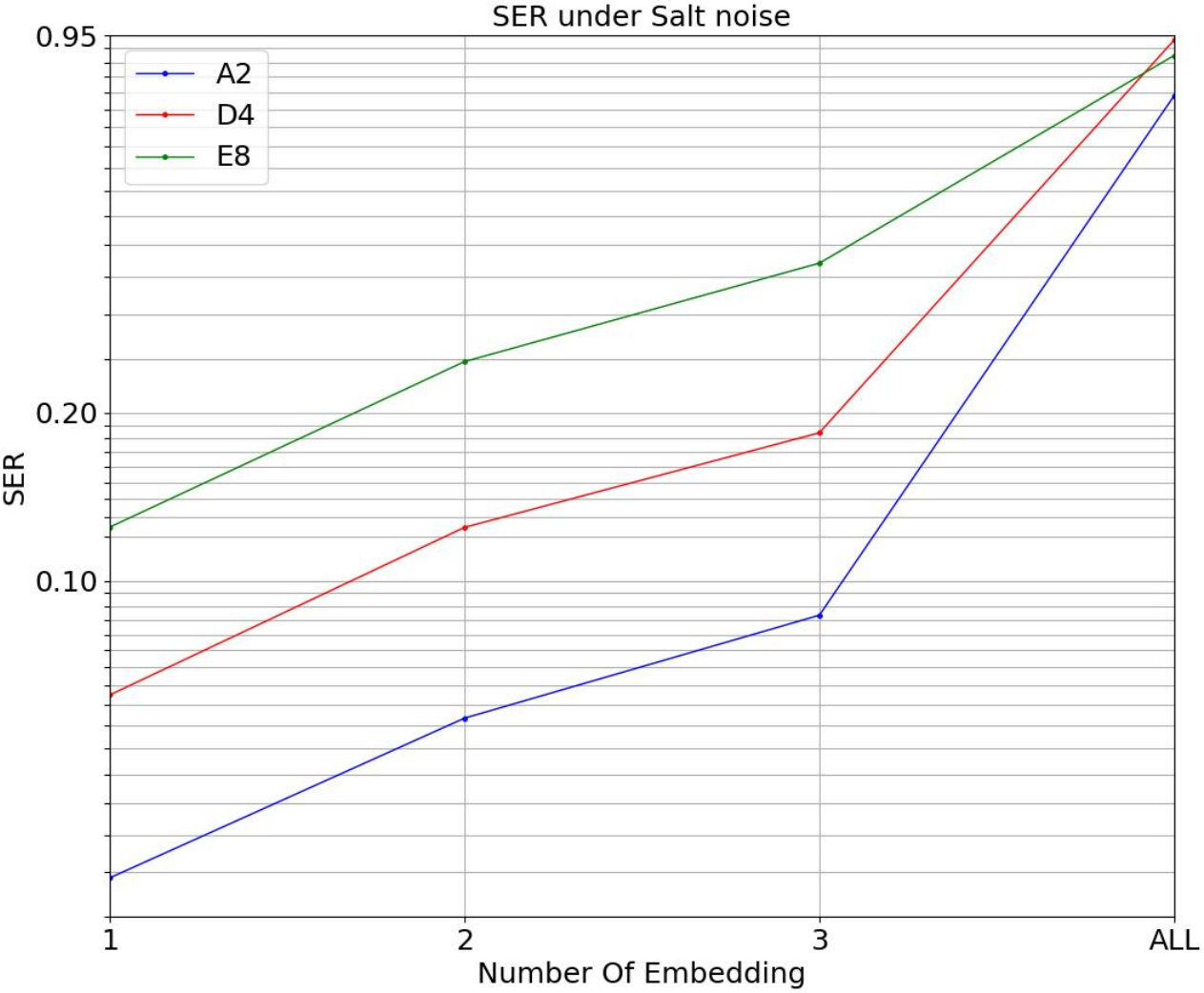}}
		\subfloat[]{\includegraphics[width=0.2\linewidth]{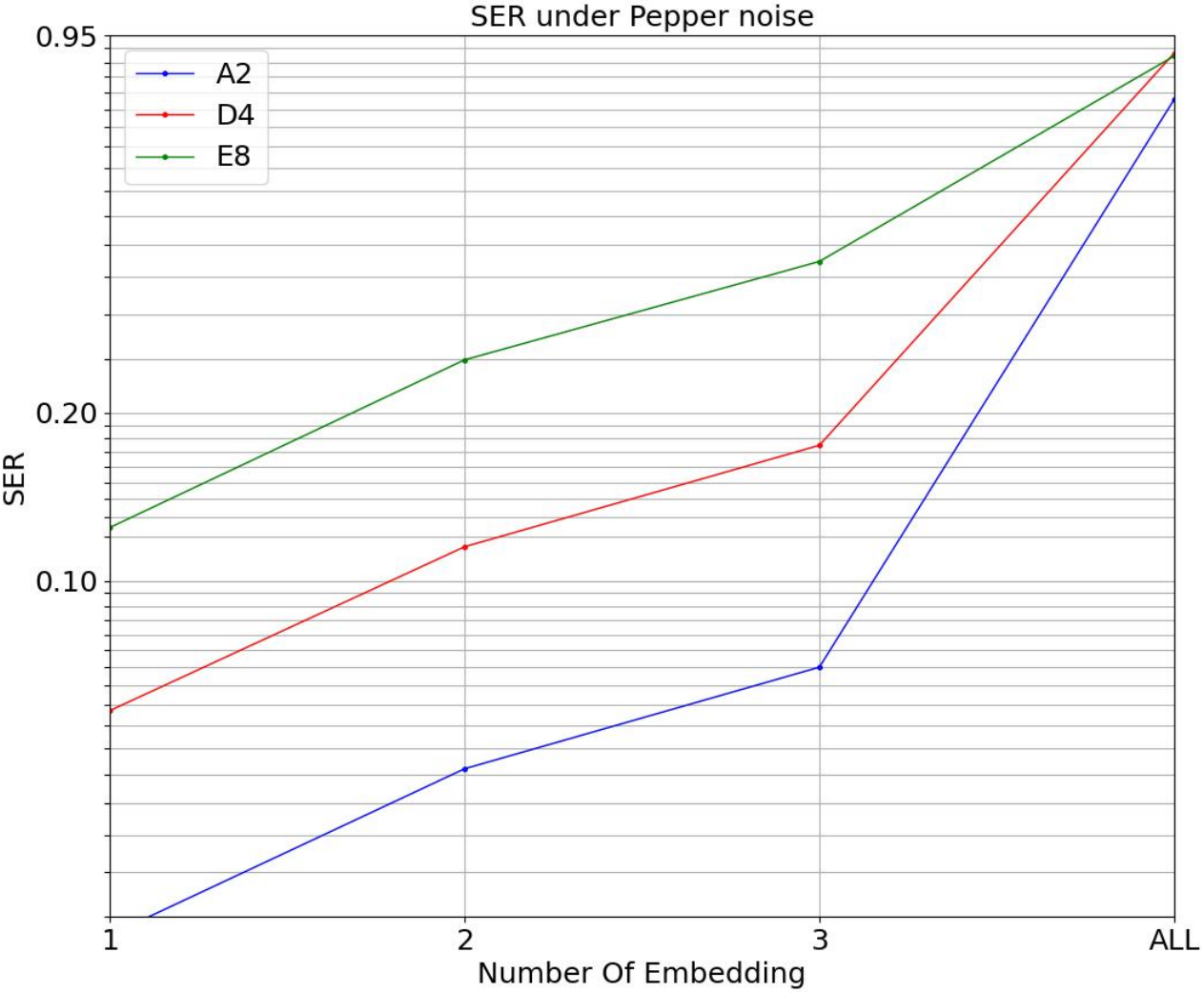}}
		\subfloat[]{\includegraphics[width=0.2\linewidth]{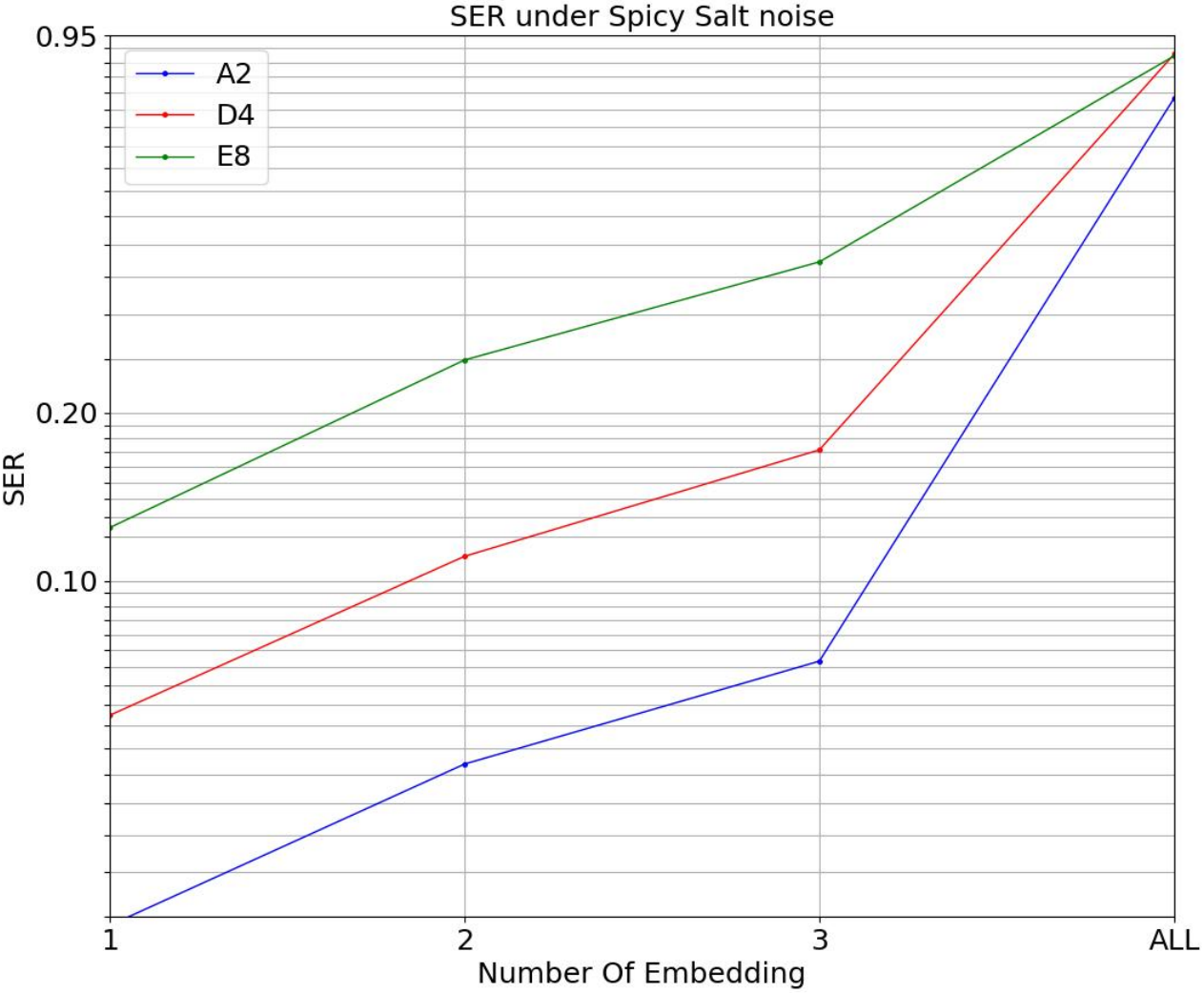}}
		\subfloat[]{\includegraphics[width=0.2\linewidth]{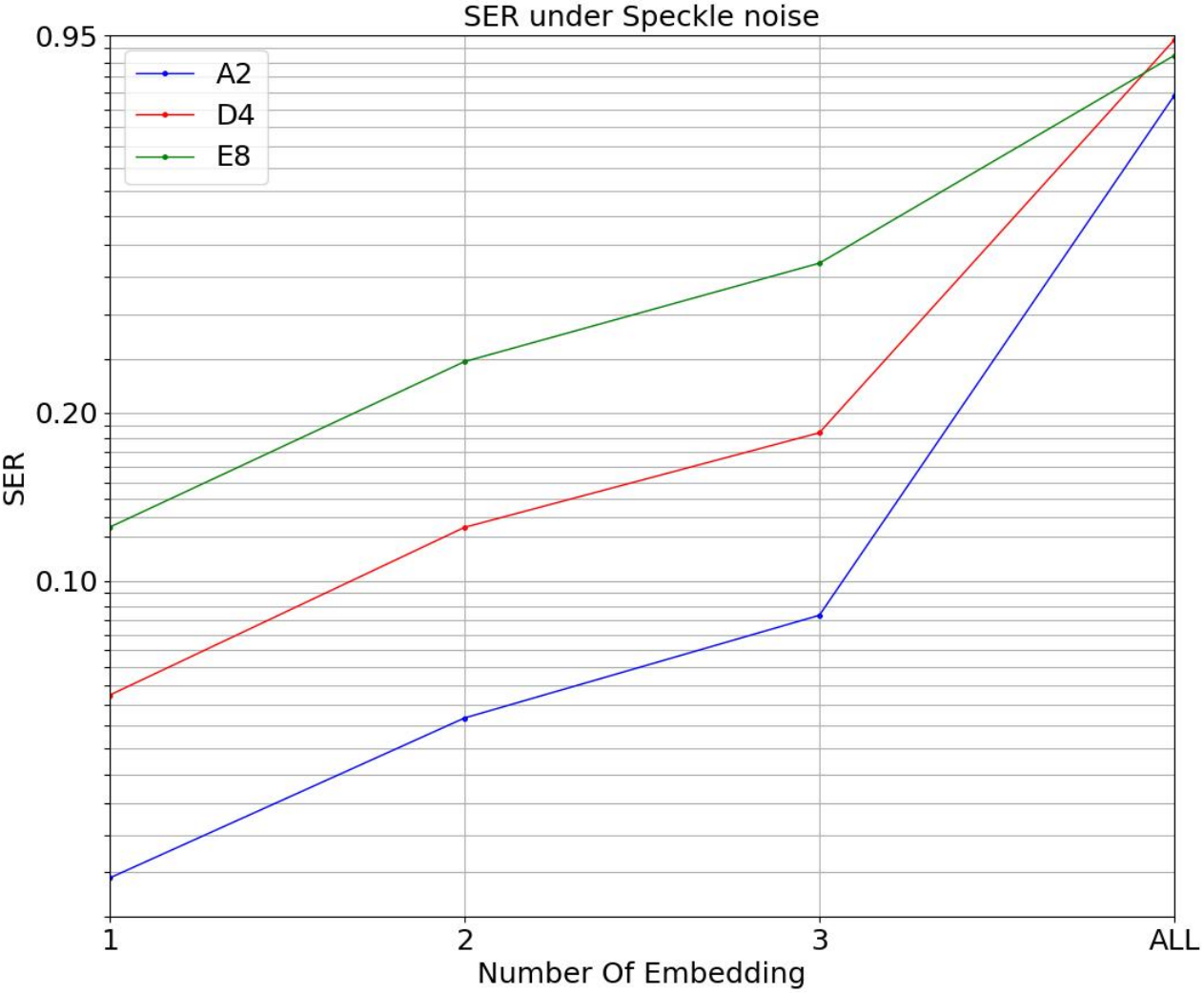}}
		\caption{The SER performance of CA-QIM over different types of noise attacks. (a) Gaussian noise (b) Salt noise (c) Pepper noise (d)Spicy Salt noise (e) Speckle noise.}
		\label{SER_fig}
	\end{figure*}

The proposed method mainly modifies the labeling process, and its robustness is equivalent to the original QIM. To evaluate the robustness, we introduce five types of noise into the embedded images and extract the messages. Since CAMD-QIM, which moves the host signals near the boundaries of the Voronoi region, exhibits poor robustness, we only test CA-QIM in this section using Symbol Error Rate (SER). We recorded the average SER in four cases: embedding one message, two messages, three messages, and full embedding under $A_2$, $D_4$, and $E_8$, and the results are displayed in Figure \ref{SER_fig}.

From Figure \ref{SER_fig}, we can make the following observations:

\begin{enumerate}
	\item In any type of noise, the SER under $A_2$ is lower than that under other lattice bases.
	
\item	As the number of embedded messages increases, the SER also increases continuously. Once the SER reaches a certain value, the final decoding effect is not ideal. Therefore, while controlling the SER, it is recommended to use the $A_2$ lattice and minimize the number of embedded messages.
\end{enumerate}

In normal QIM, the symbol error rate typically decreases as the lattice dimension increases. However, in this experiment, the results are opposite because the image pixels in the experiment are integers. After the inverse Discrete Cosine Transform (DCT), larger lattice dimensions lead to greater pixel changes, and after rounding, the deviation becomes larger.
	
	\section{Conclusions}
	\label{Conclusions}
In conclusion, this paper introduced an enhanced version of the Quantization Index Modulation (QIM) watermarking technique, incorporating a novel labeling scheme. The proposed method effectively reduces embedding distortion by adapting the codebook of host signals and messages, particularly benefiting non-uniformly distributed messages. Through extensive simulations using typical image datasets, the superiority of our approach has been demonstrated.
The research presented in this paper highlights the potential of combining QIM watermarking with big data techniques, offering a promising direction for improving the efficiency of watermarking systems. By leveraging data-driven approaches and exploiting statistical properties, we can further enhance the performance and effectiveness of image watermarking methods.

Future work in this area may involve exploring additional data-driven techniques and optimizing the proposed labeling scheme to address different challenges and scenarios. Continued research in image watermarking and its integration with big data analytics holds significant potential for advancing the field and enabling robust and efficient protection of digital media.
	
%
%
	
	\bibliographystyle{IEEEtranMine}
	\bibliography{CAQIM}
\end{document}